\documentstyle[12pt,preprint,aps,floats]{revtex}

\tighten

\input epsf.tex

\def\APJ#1#2#3{Ap. J. {\bf #1}, #2 (#3)}
\def\AstroP#1#2#3{Astropart. Phys. {\bf #1}, #2 (#3)}

\def\NIM#1#2#3{Nucl. Instrum. Meth. {\bf A#1}, #2 (#3)}

\def\NPPS#1#2#3{Nucl. Phys. Proc. Suppl. {\bf #1}, #2 (#3)}

\def\PRD#1#2#3{Phys. Rev. {\bf D#1}, #2 (#3)}

\def\journal#1#2#3#4{#1 {\bf #2}, #3 (#4)}
\newcommand{\postscript}[2]{\setlength{\epsfxsize}{#2\hsize}
   \centerline{\epsfbox{#1}}}
\newcommand{\sign}{\:\!\text{sgn}\:\!}
\newcommand{\mgaugino}{M_{1/2}}

\newcommand{\mgut}{M_{\text{GUT}}}

\newcommand{\tb}{\tan\beta}

\newcommand{\ifb}{\text{fb}^{-1}}

\newcommand{\gev}{\text{GeV}}
\newcommand{\tev}{\text{TeV}}
\newcommand{\cm}{\text{cm}}
\newcommand{\km}{\text{km}}
\newcommand{\s}{\text{s}}
\newcommand{\yr}{\text{yr}}
\newcommand{\sr}{\text{sr}}
\newcommand{\ethr}{E_{\text{thr}}}
\newcommand{\eopt}{E_{\text{opt}}}
\newcommand{\mchi}{m_{\chi}}
\newcommand{\Omegachi}{\Omega_{\chi}}
\newcommand{\be}{\begin{equation}}
\newcommand{\ee}{\end{equation}}
\newcommand{\etal}{{\em et al.}}

\newcommand{\met}{\rlap{\,/}E_T}

\begin{document}

\draft

\renewcommand{\thefootnote}{\fnsymbol{footnote}}
\setcounter{footnote}{0}

\preprint{
\noindent
\hfill
\begin{minipage}[t]{3in}
\begin{flushright}
IASSNS--HEP--00--55\\
FERMILAB--PUB--00/171--T\\
astro-ph/0008115\\
\end{flushright}
\end{minipage}
}

\title{
\vskip 0.5in
Prospects for Indirect Detection\\
of Neutralino Dark Matter}

\author{
Jonathan L.~Feng$^a$, Konstantin T.~Matchev$^b$, and 
Frank Wilczek$^a$
\vskip 0.2in
}

\address{
  ${}^{a}$
  School of Natural Sciences, Institute for Advanced Study\\
  Princeton, NJ 08540, U.S.A. 
\vskip 0.1in
  ${}^{b}$
  Theoretical Physics Department,
  Fermi National Accelerator Laboratory\\
  Batavia, IL 60510, U.S.A.}

\maketitle

\begin{abstract}
Dark matter candidates arising in models of particle physics
incorporating weak scale supersymmetry may produce detectable signals
through their annihilation into neutrinos, photons, or positrons.  A
large number of relevant experiments are planned or underway.  The
`logically possible' parameter space is unwieldy.  By working in the
framework of minimal supergravity, we can survey the implications of
the experiments for each other, as well as for direct searches,
collider searches, low-energy experiments, and naturalness in a
transparent fashion.  We find that a wide variety of experiments
provide interesting probes.  Particularly promising signals arise in
the mixed gaugino-Higgsino region.  This region is favored by
low-energy particle physics constraints and arises naturally from
minimal supergravity due to the focus point mechanism.  Indirect dark
matter searches and traditional particle searches are highly
complementary.  In cosmologically preferred models, if there are
charged superpartners with masses below 250 GeV, then some signature
of supersymmetry must appear {\em before} the LHC begins operation.
\end{abstract}


\newpage

\renewcommand{\thefootnote}{\arabic{footnote}}
\setcounter{footnote}{0}

\section{Introduction}
\label{sec:introduction}

It is now well established that luminous matter makes up only a small
fraction of the mass of the observed universe.  The evidence for dark
matter is both astrophysical and cosmological~\cite{DMreviews}.  Such
evidence requires only that dark matter is gravitationally
interacting.  However, additional constraints, especially the success
of light-element cosmonucleosynthesis calculations, strongly disfavor
the possibility that dark matter is composed solely of
baryons~\cite{Freese:1999sh}, and so some form of matter foreign to
our everyday world is required. The dark matter problem is therefore
also an important problem for particle physics, as particle physics
both suggests promising possibilities and imposes stringent
constraints.

Neutralinos are well-motivated candidates to provide much or all of
the non-baryonic dark matter.  An effectively stable particle is a
generic component of models with weak-scale supersymmetry.  This
particle is the lightest supersymmetric particle (LSP), and is
typically the neutralino, a mixture of the superpartners of Higgs and
electroweak gauge bosons.  Particle physics considerations alone
require the neutralino to be electrically neutral, effectively stable
(assuming $R$-parity conservation, which is also motivated by the need
to forbid too-rapid proton decay), and weakly interacting, with mass
of order 100 GeV (required, as we shall quantify below, if
supersymmetry naturally protects the electroweak scale from large
radiative corrections). Remarkably, these properties are consistent
with the possibility that the thermal relic density of neutralinos
makes up most of the missing mass of the
universe~\cite{Goldberg:1983nd,Ellis:1983wd}.

Unfortunately, these properties also guarantee that neutralinos are
practically impossible to observe in collider experiments directly.
They pass through collider detectors without interacting. Existing
bounds on neutralinos therefore rely on model-dependent correlations
between their properties and those of other supersymmetric
particles. At present, in minimal supergravity, the LEP experiments
constrain $m_{\chi} \agt 40~\gev$~\cite{LEPbound}. In the next several
years, Run II of the Tevatron at Fermilab and, eventually, the Large
Hadron Collider at CERN will provide more powerful collider probes.

If neutralinos make up a significant portion of the halo dark matter,
many additional avenues for their detection open up.  They may deposit
energy as they scatter off nuclei in detectors.  We have investigated
the prospects for direct detection in a companion article~\cite{FMW},
where we emphasized the importance and promise of a mixed
gaugino-Higgsino regime, previously neglected.  Here we will study the
possibility of detecting neutralinos indirectly by looking for
evidence of their annihilation~\cite{indirectreviews}.  In the next
five years, an astounding array of experiments will be sensitive to
the various potential neutralino annihilation products. These include
under-ice and underwater neutrino telescopes (AMANDA, NESTOR,
ANTARES), atmospheric Cherenkov telescopes (STACEE, CELESTE, ARGO-YBJ,
MAGIC, HESS, CANGAROO, VERITAS), space-based $\gamma$ ray detectors
(AGILE, AMS/$\gamma$, GLAST), and anti-matter/anti-particle
experiments (PAMELA, AMS).  In many cases, these experiments will
improve current sensitivities by several orders of magnitude.

In this paper we evaluate the prospects for neutralino dark matter
discovery through indirect detection.  The neutralino signals depend
on many unknown parameters.  At the same time, an abundance of
theoretical and experimental information from particle physics can be
brought to bear.  The implications of traditional particle physics
searches for dark matter searches, and {\it vice versa}, are already
significant, and promise to become much stronger over the next few
years.  One of our main conclusions is that in a class of particle
physics models favored by current particle physics constraints,
astrophysical signals are especially enhanced.

Previous discussions of indirect neutralino detection fall rather
sharply into two schools. Several previous works are based on specific
high energy models~\cite{Faraggi:2000iu,Corsetti:2000yq}, including a
number in the framework of minimal supergravity~\cite{Giudice:1989vs,%
Gelmini:1991je,Gandhi:1994ce,Diehl:1995ff,Corsetti:2000ma}.  As we
will recall below, minimal supergravity incorporates several desirable
features, including the radiative breaking of electroweak symmetry and
the possibility, suggested by gauge coupling
unification~\cite{Dimopoulos:1981yj}, of perturbative extrapolation to
large energy scales.  Previous studies in minimal supergravity have
concluded that only Bino-like dark matter is allowed by particle
physics constraints.  Such dark matter necessarily implies highly
suppressed dark matter signals, as we will see.  These studies, and
their somber conclusions, have been criticized as products of overly
restrictive particle physics assumptions~\cite{Bergstrom:1996cz}.
Recently we have argued more specifically that, even in minimal
models, these studies failed to examine a very well-motivated regime
of parameters, and that for this reason their conclusions are overly
pessimistic~\cite{FMW}.

At the other extreme, several studies scan over a large set of
weak-scale supersymmetry parameters and consider values for these
parameters as large as 50 TeV. (See, for example,
Refs.~\cite{Bergstrom:1996cz,Bergstrom:1997kp,Bergstrom:1998xh,%
Buckley,Baltz:2000ra}.)  These studies, and others, bring a high level
of sophistication to the evaluation of astrophysical effects on dark
matter signals.  In this regard, we will have nothing to add, but we
will incorporate many of the most accurate recent results in our
study.

{}From a particle physics perspective, this second group of studies is
impressively general, but this generality is achieved at a cost.  For
example, extrapolating a given set of weak-scale parameters to higher
scales, one may encounter such diseases as Landau poles or charge- or
color-breaking minima.  Models of this sort do not arise within a
reasonable high energy framework.  In addition, there is the practical
difficulty that the proliferation of free parameters implies that
results cannot be presented in a systematic, yet transparent, fashion.
Typically they are displayed as scatter plots after scanning over all
parameters.  It is nearly impossible, from such plots, to determine
the dependence of the signal rates on the underlying physical
parameters.  Dedicated correlation plots have been used to highlight a
few of the relations between dark matter detection experiments, but
even the most general implications of these experiments for collider
searches, electric dipole moments, anomalous magnetic moments, proton
decay, flavor violation, and other searches for supersymmetry are very
hard to discern.  Finally, it can be expected that supersymmetry
parameters, such as the $\mu$ and gaugino mass parameters, of order 50
TeV will require fine-tuning of the order of 1 part in $10^6$ to
produce the observed electroweak scale. (While it is impossible to
speak of naturalness without first specifying a mechanism of
electroweak symmetry breaking, at present all concrete models display
this feature.)  Such large fine-tuning destroys one of the main
motivations for considering supersymmetric extensions of the standard
model in the first place.  Lacking both a systematic framework and a
systematic presentation, it is impossible to see how the expectations
narrow when some naturalness criterion is imposed.

It may be impossible to satisfy both schools simultaneously.  However,
recent results suggest an appealing compromise.  As has been
emphasized in Refs.~\cite{FMM1,FMM2,FMM3}, a Bino-like LSP is {\em
not\/} a robust prediction of minimal supergravity. We find that both
Bino-like and mixed gaugino-Higgsino dark matter is possible. So
simply by considering all of minimal supergravity parameter space
carefully, as we will do here, we remove the most egregious form of
model dependence.  At the same time, by staying within the confines of
minimal supergravity we will be able to present results in an
organized and comprehensive manner, so that correlations with all
other supersymmetric signals are easily determined.  As the
experiments discussed below report results, it will be ever more
interesting to see what models are being excluded or favored.  The
framework discussed here makes this possible.

Inclusion of the new gaugino-Higgsino LSP models in minimal
supergravity is not just a formality. The region with mixed
gaugino-Higgsino LSPs is now known to be robust and natural, given an
objective definition of naturalness~\cite{FMM1,FMM2,FMM3}.  It yields
cosmologically interesting relic densities~\cite{FMW}, and is even
favored by low energy constraints such as proton decay and electric
dipole moments~\cite{FM}.  As we will see, {\em all\/} indirect
detection signals are enhanced in this region.  This lends increased
interest to indirect dark matter searches, since large --- possibly
spectacular --- rates are predicted within an attractive and simple
high energy framework.

In the following section, we review a few essential results concerning
neutralino dark matter in minimal supergravity with an emphasis on the
new gaugino-Higgsino LSP region.  We then consider each of three
promising signals in the following sections: upward-going muons from
neutrinos in Sec.~\ref{sec:neutrinos}, photons in
Sec.~\ref{sec:photons}, and positrons in Sec.~\ref{sec:positrons}.  In
Sec.~\ref{sec:comp} we compare these to direct dark matter and
traditional particle physics searches, and in
Sec.~\ref{sec:conclusions} we summarize our results.

\section{Neutralino Dark Matter and its Annihilation}
\label{sec:neutralinos}

The lightest neutralino is the LSP in many supersymmetric models.
Assuming $R$-parity conservation to prevent too-rapid proton decay
through dimension-four operators, the LSP is effectively stable, and
the neutralino is then an excellent candidate for cold dark matter.

The signals of neutralino dark matter are determined in large part by
their composition.  Neutralinos are mixtures of the superpartners of
Higgs and electroweak gauge bosons.  After electroweak symmetry
breaking, these gauge eigenstates mix through the tree-level mass
matrix

\begin{equation}
\label{neumass}
\left( \begin{array}{cccc}
M_1        &0       
&-m_Z \cos\beta\, \sin \theta_W & m_Z \sin\beta\, \sin \theta_W \\
0          &M_2     & m_Z \cos\beta\, \cos
\theta_W &-m_Z \sin\beta\, \cos
\theta_W \\
-m_Z \cos\beta\, \sin \theta_W  & m_Z \cos\beta\, \cos
\theta_W &0       &-\mu      \\
 m_Z \sin\beta\, \sin \theta_W  &-m_Z \sin\beta\, \cos
\theta_W &-\mu    &0 \end{array}
\right)
\end{equation}
in the basis $(-i\tilde{B},-i\tilde{W}^3, \tilde{H}^0_u,
\tilde{H}^0_d)$.  The weak scale parameters entering this mass matrix
are the Bino, Wino, and Higgsino mass parameters $M_1$, $M_2$, and
$\mu$, and the ratio of Higgs scalar vacuum expectation values $\tb =
\langle H^0_u \rangle / \langle H^0_d \rangle$.  The lightest
neutralino mass eigenstate is then determined by these parameters to
be some mixture
\begin{equation}
\chi \equiv \chi^0_1 = a_1 (-i \tilde{B}) + a_2 (-i \tilde{W}^3) 
+ a_3 \tilde{H}^0_u + a_4 \tilde{H}^0_d \ .
\end{equation}
We define the LSP gaugino fraction to be
\begin{equation}
R_{\chi} \equiv |a_1|^2 + |a_2|^2 \ .
\end{equation}
In the following, we refer to neutralinos with $0.9 < R_{\chi}$ as
gaugino-like, $0.1 \le R_{\chi} \le 0.9$ as mixed gaugino-Higgsino,
and $R_{\chi} < 0.1$ as Higgsino-like.

The preceding discussion is model-independent, assuming only minimal
field content.  However, the minimal supersymmetric standard model is
undoubtedly a low-energy effective theory of a more fundamental theory
defined at some higher scale, such as the grand unified theory (GUT)
or string scale.  A simple realization of this idea is the framework
of minimal supergravity, which is fully specified by the five
parameters (four continuous, one binary)
\begin{equation} 
m_0, \mgaugino, A_0, \tb, \sign (\mu) \ .  
\end{equation} 
Here, $m_0$, $\mgaugino$, and $A_0$ are the universal scalar mass,
gaugino mass, and trilinear scalar coupling.  They are assumed to
arise through supersymmetry breaking in a hidden sector at the GUT
scale $\mgut \simeq 2 \times 10^{16}~\gev$.  The hidden-sector
parameters then determine all the couplings and masses of the weak
scale Lagrangian through renormalization group evolution.  In
particular, electroweak symmetry is broken radiatively by the effects
of a large top quark Yukawa coupling, and the electroweak scale is
determined in terms of supersymmetry parameters through
\begin{equation}
\frac{1}{2} m_Z^2 = \frac{m_{H_d}^2 - m_{H_u}^2 \tan^2\beta }
{\tan^2\beta -1} - \mu^2\ .
\label{mZ}
\end{equation}
Equation~(\ref{mZ}) receives corrections beyond tree-level; in our
work we include all one-loop effects in the Higgs
potential~\cite{Pierce:1997zz}.  We also use two-loop renormalization
group equations~\cite{2loop RGEs} with one-loop threshold
corrections~\cite{Pierce:1997zz,Bagger:1995bw} and calculate all
superpartner masses to one-loop~\cite{Pierce:1997zz}. All of the
qualitative features to be described below are present, however, also
for one-loop renormalization group equations.

Minimal supergravity is, of course, by no means the most general
allowed framework.  It is worth noting, however, that the assumptions
most relevant for dark matter, namely, the universality of gaugino and
scalar masses, are motivated not only by their simplicity, but also by
concrete experimental facts. The case for gaugino mass unification is
especially powerful.  In the minimal supersymmetric standard model,
the three gauge couplings, when evolved to high scales, meet with high
precision at a point, the GUT scale~\cite{Dimopoulos:1981yj}.  If the
standard model is unified in a grand unified gauge theory, one
typically obtains also gaugino mass unification.  The unification of
couplings calculation also distinguishes $\mgut \simeq 2 \times
10^{16}~\gev$ as the natural scale for a more fundamental framework.

The unification of scalar masses is motivated by a similar, although
more speculative, argument.  Consider the mass parameter $m_{H_u}^2$,
which, from Eq.~(\ref{mZ}), plays the critical role in determining the
weak scale for all moderate and large values of $\tb$ ($\tb \agt 5$).
For theories with a universal scalar mass, it is a remarkable fact
that the renormalization group trajectories of $m_{H_u}^2$ for various
initial conditions $m_0$, when evolved to low scales, meet with high
precision at a point, the weak scale~\cite{FMM1,FMM2}.  This focusing,
which requires that the top quark mass be within $\sim 5~\gev$ of its
measured value, implies that the electroweak potential is highly
insensitive to $m_0$.  The longstanding problems of supersymmetry with
respect to CP violation, proton decay, etc. can therefore be
ameliorated without fine-tuning, simply by assuming large scalar
masses.  Although the focusing property holds more generally, its
minimal and most concrete realization is in theories with a universal
scalar mass. The assumption of a universal (and large) scalar mass is
therefore motivated by the fact that it provides a simple and elegant
solution to several well-known phenomenological problems of weak scale
supersymmetry.

\begin{figure}[!t]
\postscript{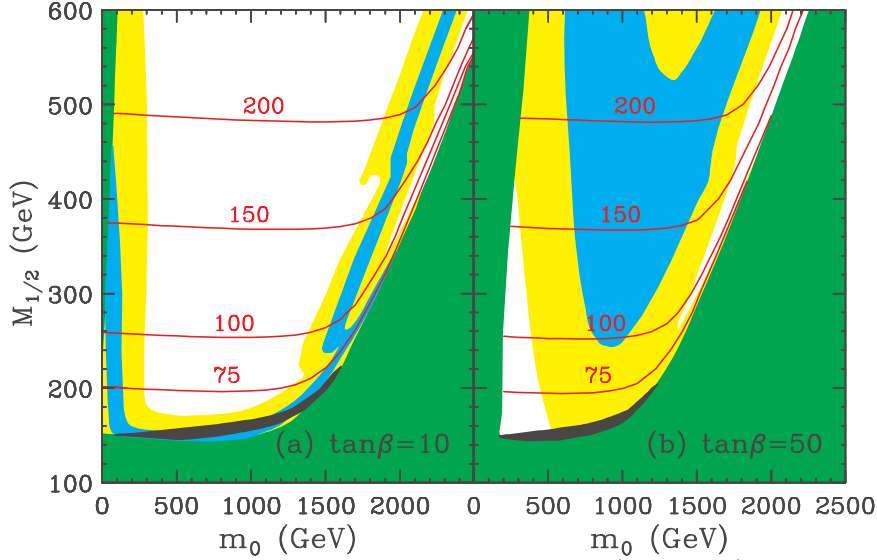}{0.70}
\caption{Contours of constant LSP mass $m_\chi$ in GeV in the $(m_0,
\mgaugino)$ plane for $A_0=0$, $\mu>0$, $m_t = 174~\gev$, and two
representative values of $\tb$. The green shaded regions are excluded
by the requirement that the LSP be neutral (left) and by the chargino
mass limit of 95 GeV (bottom and right).  We have also delineated the
regions with potentially interesting values of the LSP relic
abundance: $0.025\le \Omegachi h^2 \le 1$ (yellow) and $0.1 \le
\Omegachi h^2 \le 0.3$ (light blue). In the black region, $|2 \mchi -
m_h| < 5~\gev$, and neutralino annihilation is enhanced by a Higgs
resonance.}
\label{fig:mlsp}
\end{figure}

\begin{figure}[!th]
\postscript{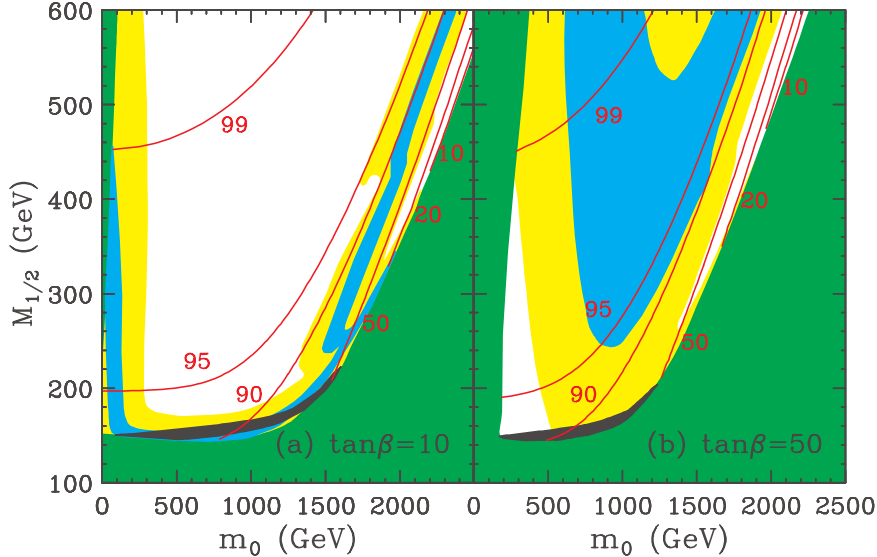}{0.70}
\caption{Contours of constant gaugino fraction $R_{\chi}$ in percent,
for the same values of the parameters as in Fig.~\ref{fig:mlsp}.}
\label{fig:gfrac}
\end{figure}

Given these motivations for minimal supergravity, we now consider
their implications for neutralino dark matter.  Gaugino mass
unification implies $M_1 \simeq M_2 / 2 \simeq 0.4 \mgaugino$.  Dark
matter is therefore never Wino-like,\footnote{Wino-like LSPs exist in
other frameworks, but typically they annihilate far too quickly to be
cosmologically relevant~\cite{Mizuta:1993ja}.  Interesting relic
densities are possible, however, if there is some mechanism of late
production~\cite{Moroi:2000zb}.} and in fact, throughout parameter
space, $|a_2|^2 < 0.07$. Additional insights follow from re-writing
Eq.~(\ref{mZ}) in terms of GUT scale parameters.  For $A_0=0$ and
$\tb=10$, for example,
\begin{equation}
\frac{1}{2} m_Z^2 \approx -0.02\; m_0^2 + 0.7\; \mgaugino^2 - \mu^2\ ,
\label{mz2}
\end{equation}
where the numerical coefficients of the first two terms vary
fractionally by ${\cal O}(10\%)$ in the $(m_0, \mgaugino)$
plane~\cite{FMM1,FMM2}.  The coefficient of $m_0^2$ is highly
suppressed~\cite{Barbieri:1988fn,Carena:1994bv}.  This is another
expression of the focusing behavior discussed above, and implies that
multi-TeV values of $m_0$ do not involve significant large
fine-tuning.  The coefficient is also negative.  For fixed
$\mgaugino$, as $m_0$ increases, $|\mu|$ decreases, and the LSP
becomes increasingly Higgsino-like. This is important, because even a
10\% Higgsino admixture drastically affects the phenomenology.  In
Figs.~\ref{fig:mlsp} and \ref{fig:gfrac} we show the LSP mass and
gaugino fraction in the $(m_0, \mgaugino)$ plane. For large $m_0 \agt
1~\tev$, we find a region, previously ignored, where the LSP has
a significant Higgsino component. The green shaded
regions are excluded by the requirement that the LSP be neutral
(top left) and by the chargino mass limit of 95 GeV (bottom and right).


In Figs.~\ref{fig:mlsp} and \ref{fig:gfrac}, we also indicate the
regions of cosmologically interesting relic densities; see
Ref.~\cite{FMW} for details.  A generous range is $0.025\le \Omegachi
h^2 \le 1$, where the lower bound is the requirement that neutralino
dark matter explain galactic rotation curves and the upper bound
follows from the lifetime of the universe.  Above this shaded region,
$\Omega h^2 > 1$; below, $\Omega h^2 < 0.025$.  The range $0.1 \le
\Omegachi h^2 \le 0.3$ is most preferred by current limits.  Our relic
density calculation is not trustworthy in the black region, where
there is a Higgs scalar resonance, and very near to the left and right
borders of the excluded region, where co-annihilation is
important~\cite{Mizuta:1993qp,Ellis:1998kh,Ellis:2000mm,Gomez:2000sj}.
In the bulk of parameter space, however, these effects are negligible.
For all $\tb$, cosmologically interesting densities are possible for
$m_0 \agt 1~\tev$.  For $\tb=10$, the cosmologically preferred region
contains gaugino-Higgsino dark matter.

In contrast to $m_0$, the parameters $\mgaugino$ and $\mu$ enter
Eq.~(\ref{mz2}) with ${\cal O}(1)$ coefficients.  Naturalness
therefore requires that the LSP mass (and, in fact, the masses of all
four neutralinos and both charginos) should not be too far above the
electroweak scale. While in principle it is possible that in some
fundamental framework $\mgaugino$ and $\mu$ are correlated precisely
in a way that allows both parameters to be large without fine-tuning
(a possibility considered in Ref.~\cite{Chankowski:1999xv}), no such
framework has been found to date.  Barring such a possibility, extreme
values such as $\mgaugino, \mu \sim 50~\tev$ require a fine-tuning of
1 part in $10^6$ and destroy one of the prime motivations for
weak-scale supersymmetry.  We therefore regard such large values as
highly disfavored, and we will focus on neutralino masses of order 100
GeV.

\begin{figure}[t]
\begin{minipage}[t]{.32\textwidth}
\postscript{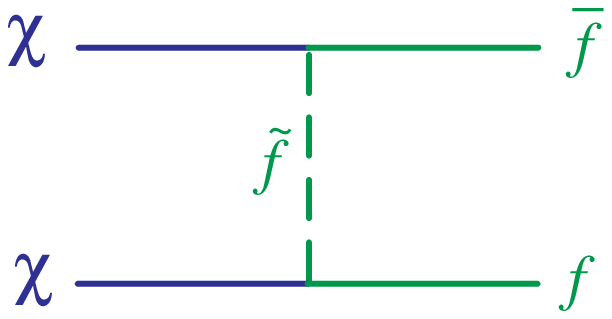}{0.89}
\end{minipage}
\begin{minipage}[t]{.32\textwidth}
\postscript{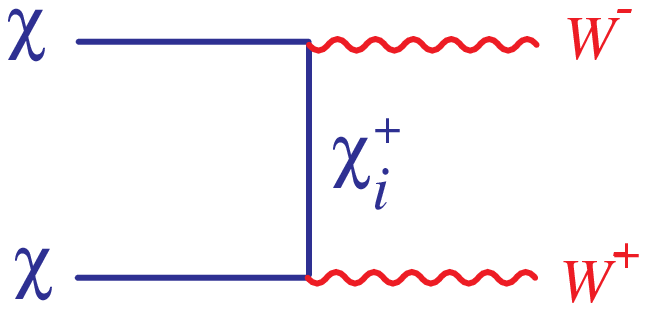}{0.93}
\end{minipage}
\begin{minipage}[t]{.32\textwidth}
\postscript{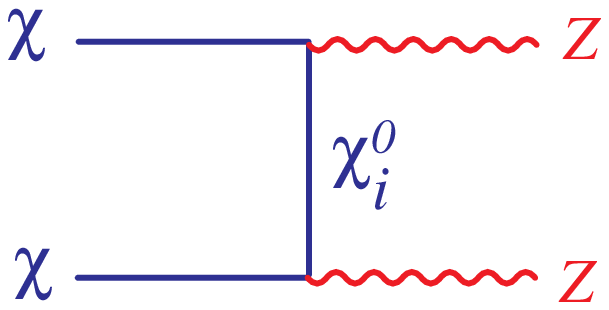}{0.89}
\end{minipage}
\caption{Three leading neutralino annihilation channels.}
\label{fig:annih}
\end{figure}

\begin{figure}[t]
\postscript{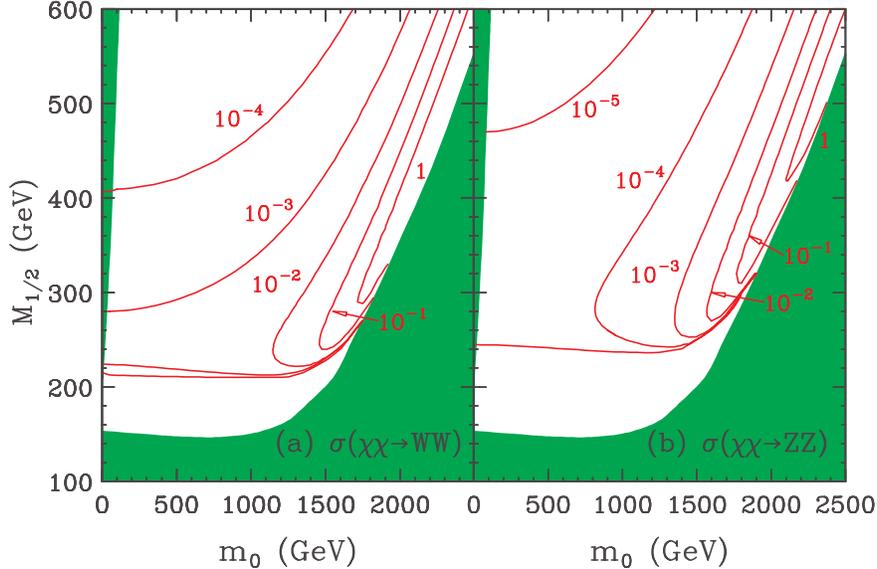}{0.70}
\caption{Contours of constant $\sigma v$ in pb for (a) $\chi \chi \to
WW$ and (b) $\chi \chi \to ZZ$. We fix $A_0=0$, $\mu>0$, $m_t =
174~\gev$, and $\tb=10$. }
\label{fig:gaugeboson}
\end{figure}

Neutralinos annihilate through a variety of channels.  The three
leading processes are shown in Fig.~\ref{fig:annih}.  (Note that
co-annihilation, while potentially important in determining relic
densities in the early universe, is negligible now.)  Annihilation
into gauge bosons is of particular importance, as these processes lead
to more energetic and striking signals.  The $WW$ cross section relies
on $W \chi \chi_i^{\pm}$ interactions.  The only such couplings
allowed by gauge invariance are $W \tilde{H}^0 \tilde{H}^{\pm}$ and $W
\tilde{W}^0 \tilde{W}^{\pm}$.  However, as noted above, gaugino mass
unification implies that the Wino content of the LSP is always
negligible. A large $WW$ cross section is therefore possible only when
the LSP has a significant Higgsino component.  The same conclusion
holds for the $ZZ$ process, where the $Z \chi \chi_i^0$ interaction is
possible only through $Z \tilde{H}^0 \tilde{H}^0$ couplings.  In
Fig.~\ref{fig:gaugeboson}, we see that the annihilation cross sections
for $\chi \chi \to WW$ and $\chi \chi \to ZZ$ are indeed highly
suppressed in regions with Bino-like LSPs, but are enhanced by three
to four orders of magnitude in regions with mixed gaugino-Higgsino
dark matter.  As we will see, this region, favored by low energy
constraints, will be the most promising for all indirect signals.

Before closing this section, we note several features of
Figs.~\ref{fig:mlsp} and \ref{fig:gfrac} that will also apply to many
of the following figures.  Unless otherwise noted, we present results
for $A_0 = 0$, $\mu >0$, $m_t = 174~\gev$, and representative values
of $\tb$ as indicated.  $A_0$ governs the left-right mixing of
scalars, and does not enter the neutralino sector.  It is therefore
largely irrelevant, especially in the regions of parameter space with
observable signals --- where, as we will see, the scalars are heavy
and decoupled.\footnote{For $A_0 = 0$, the 1999 LEP bound on the mass
of the lightest CP-even Higgs boson $m_h > 107.7 ~\gev$ barely
constrains the parameter space shown in our figures, while the 2000
limit $m_h > 113.3~\gev$ excludes the region with $m_0\alt 1$ TeV and
$M_{1/2}\alt 300$ GeV.  However, the Higgs search is highly sensitive
to the $A_0$ parameter.  For other values of $A_0$, the current
constraints may be evaded with negligible impact on our predictions
for indirect dark matter detection.  For this reason, we do not
include the Higgs mass constraints in the following analysis.}
(Besides, the most important trilinear coupling, $A_t$,
has a weak-scale fixed point, and so is only weakly sensitive to
$A_0$.)  Our dark matter results are rather insensitive to the sign of
$\mu$, but the choice $\mu>0$ is motivated by the constraint from $B
\to X_s \gamma$ (see, e.g. \cite{Baer:1997kv}).
Finally, perturbativity of Yukawa couplings limits
$\tb$ to the range $1 \alt \tb \alt 60$.  Low values of $\tb \alt 3$
are now being excluded by the LEP Higgs search.  In the remaining
interval, models with moderate and high values may have qualitatively
different behavior, as processes proportional to down-type Yukawa
couplings are enhanced by $\tb$.  We therefore typically present
results for two representative values, one in each range. Plots of
many other quantities, including all physical slepton, squark, and
Higgs masses in the $(m_0, \mgaugino)$ plane, including the high $m_0$
region, may be found in Refs.~\cite{FMM1,FMM2,FMM3}.

Finally, in the following three sections, we will {\em not} include
the effects of variations in thermal relic density in our signal
rates, but rather assume, for concreteness, a fixed local neutralino
density.  The results are then more transparent, and are applicable to
general scenarios, such as those in which a late source of neutralino
production is present.  Of course, in the simplest scenario, models
with $\Omegachi h^2 > 1$ are excluded, and those with an
under-abundance of neutralinos are disfavored and imply suppressed or
negligible dark matter signals. This should be kept in mind in the
following sections.  In Secs.~\ref{sec:comp} and
\ref{sec:conclusions}, we will combine all these considerations, and
focus on the most preferred regions.

\section{Neutrinos}
\label{sec:neutrinos}

When neutralinos pass through astrophysical objects, they may be
slowed below escape velocity by elastic scattering.  Once captured,
they then settle to the center, where their densities and annihilation
rates are greatly enhanced.  While most of their annihilation products
are immediately absorbed, neutrinos are not.  High energy neutrinos
from the cores of the Earth~\cite{Freese:1986qw,Krauss:1986aa,%
Gaisser:1986ha,Gould:1989eq} and Sun~\cite{Gaisser:1986ha,%
Press:1985ug,Silk:1985ax,Srednicki:1987vj,Hagelin:1986gv,%
Ng:1987qt,Ellis:1988sh} are therefore promising signals for indirect
dark matter detection.

The formalism for calculating neutrino fluxes from dark matter
annihilation is well developed. (See Ref.~\cite{Jungman:1996df} for a
review.)  The neutrino flux depends first and foremost on the
neutralino density, which is governed by the competing processes of
gravitational capture and neutralino annihilation.  If $N$ is the
number of neutralinos in the Earth or Sun, $\dot{N} = C - A N^2$,
where $C$ is the capture rate and $A$ is the total annihilation cross
section times relative velocity per volume.  The present neutralino
annihilation rate is then
\begin{equation}
\Gamma_A = \frac{1}{2} A N^2 = \frac{1}{2} C \tanh^2(\sqrt{CA}\,
t_{\odot}) \ ,
\end{equation}
where $t_{\odot} \approx 4.5$ Gyr is the age of the solar system.

Captured neutralinos then annihilate through the processes of
Fig.~(\ref{fig:annih}).  As $\chi \chi \to f \bar{f}$ is
helicity-suppressed, neutrinos are produced only in the decays of
primary annihilation products.  Typical neutrino energies are then
$E_{\nu} \sim \frac{1}{2}\mchi$ to $\frac{1}{3}\mchi$, with the most
energetic spectra resulting from $WW$, $ZZ$, and, to a lesser extent,
$\tau\bar{\tau}$.  After propagating to the Earth's surface, neutrinos
are detected through their charged-current interactions.  The most
promising signal is from upward-going muon neutrinos that convert to
muons in the surrounding rock, water, or ice, producing through-going
muons in detectors.  The detection rate for such neutrinos is greatly
enhanced for high energy neutrinos, as both the charged-current cross
section and the muon range are proportional to $E_{\nu}$.

The calculation of muon fluxes from neutralino annihilation in the
Earth and Sun is on reasonably firm footing, as it depends only on the
local dark matter density and is insensitive to details of halo
modeling.\footnote{It has been suggested that muon fluxes may be
enhanced, by up to two orders of magnitude, due to capture of
neutralinos in highly eccentric solar system
orbits~\cite{Bergstrom:1999tk}. The magnitude of the enhancement
depends on details of the neutralino parameters and involves
astrophysical issues still under debate~\cite{Gould:1999je}. We have
not included it here.}  Nevertheless, the calculation is involved,
primarily as a result of complications in evaluating capture
rates~\cite{Gould:1987ju,Gould:1987ir,Gould:1988ww} and, in the case
of the Sun, in propagating the neutrinos from the core to the
surface~\cite{Gaisser:1986ha,Ritz:1988mh,Jungman:1995jr}.
Here we make use of the procedure of
Refs.~\cite{Jungman:1996df,Kamionkowski:1991nj}.  For other analyses,
see Refs.~\cite{Bottino:1991dy,Bottino:1995xp,Berezinsky:1996ga,%
Bergstrom:1997kp}, as well as those motivated by the Tevatron
$e^+e^-\gamma\gamma$ event~\cite{Freese:1997kq,Bottino:1997hv} and by
the DAMA annual modulation
signal~\cite{Bottino:1999vw,Bottino:2000gc}.

\begin{figure}[!t]
\postscript{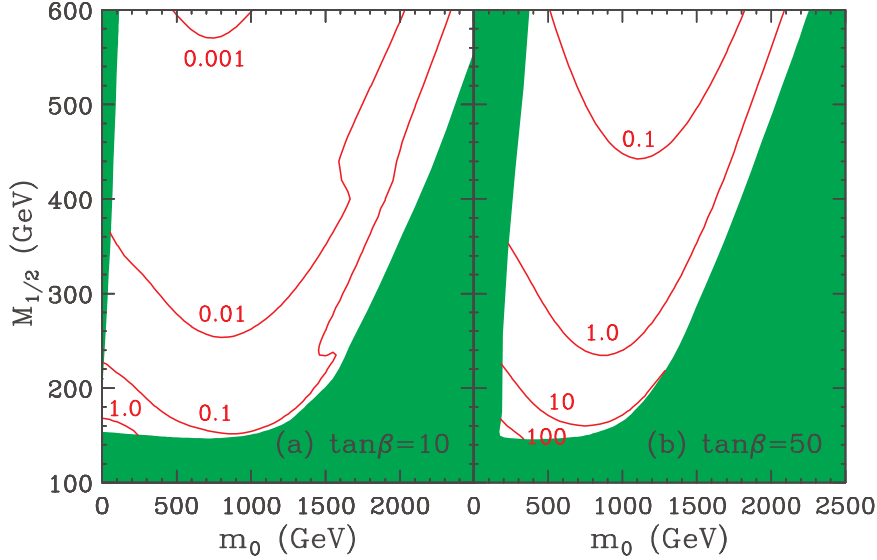}{0.70}
\caption{The filling parameter $\sqrt{CA}\, t_{\odot}$ for the Earth.
We assume neutralino velocity dispersion $\bar{v} = 270~\km/\s$ and a
local density of $\rho_0 = 0.3~\gev/\cm^3$.}
\label{fig:t_tauE}
\end{figure}

\begin{figure}[h]
\postscript{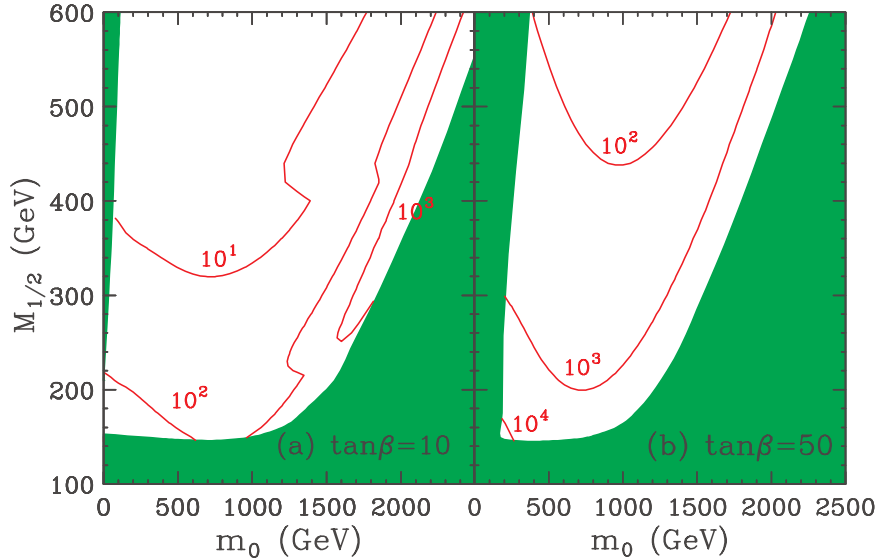}{0.70}
\caption{As in Fig.~\ref{fig:t_tauE}, but for the Sun.}
\label{fig:t_tauS}
\end{figure}

The `filling parameters' $\sqrt{CA}\, t_{\odot}$ for the Earth and Sun
are given in Figs.~\ref{fig:t_tauE} and \ref{fig:t_tauS}.  For the
Sun, $\sqrt{CA}\,t_{\odot} \gg 1$ for all supersymmetry parameters.
The neutralino density has therefore reached equilibrium, and the
annihilation rate is at full strength, with $\Gamma_A \approx C/2$.
For the Earth, however, typically $\sqrt{CA}\, t_{\odot} \ll 1$, and
the annihilation rate is $\Gamma_A \approx \frac{1}{2} C^2 A\,
t_{\odot}^2$ and far from maximal.  As we will see, this plays an
important part in reducing the Earth's signal below the Sun's.

\begin{figure}[!t]
\postscript{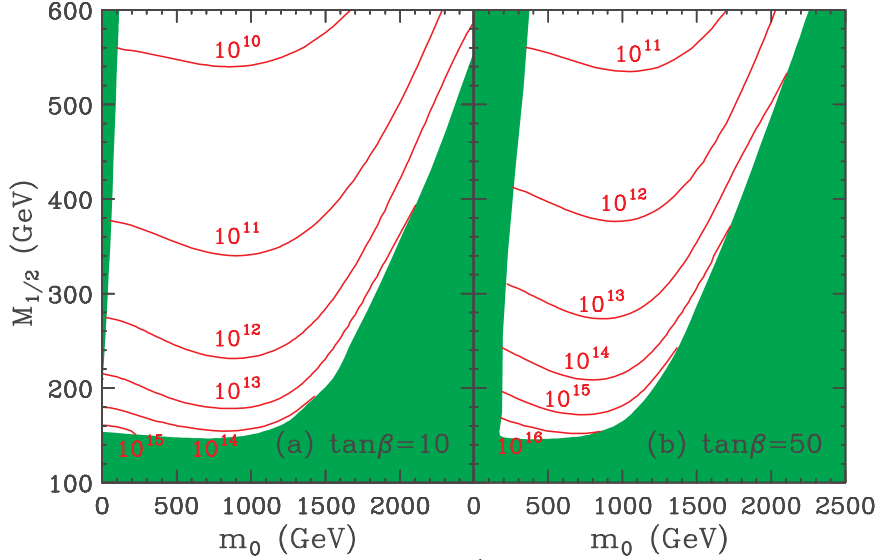}{0.70}
\caption{The capture rate of neutralinos $C$ in $\s^{-1}$ in the
center of the Earth (for $\bar{v} = 270~\km/\s$ and
$\rho_0 = 0.3~\gev/\cm^3$). }
\label{fig:captureE}
\end{figure}

\begin{figure}[h]
\postscript{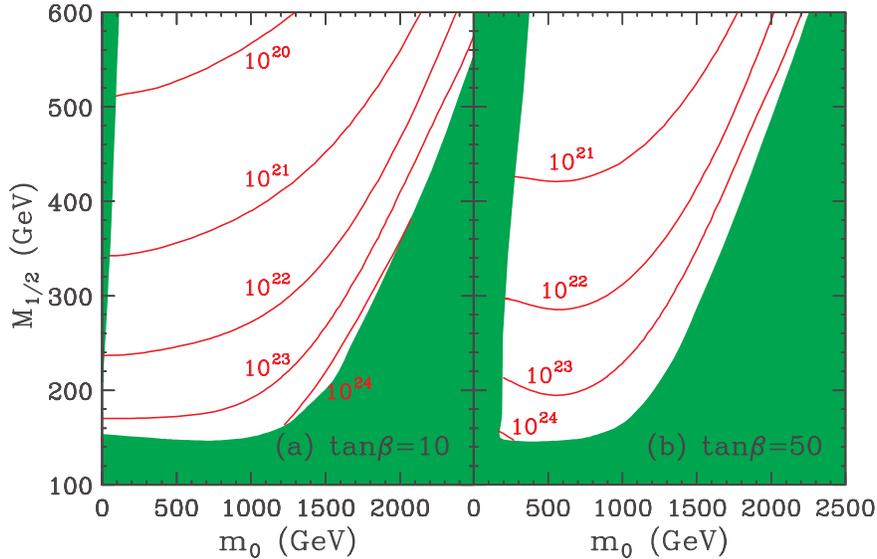}{0.70}
\caption{Same as in Fig.~\ref{fig:captureE}, but for the Sun.}
\label{fig:captureS}
\end{figure}

The other major ingredient in the muon flux computation is the
estimate of the neutralino capture rate $C$, which is shown in
Figs.~\ref{fig:captureE} and \ref{fig:captureS} for the Earth and Sun,
respectively. The elemental compositions of the Earth and Sun are
given in Ref.~\cite{Gelmini:1991je}.  A quick comparison of
Figs.~\ref{fig:captureE} and \ref{fig:captureS} reveals that, not
surprisingly, a large astrophysical body like the Sun is much more
efficient in trapping neutralinos.  The Earth's capture rate is,
however, enhanced by the iron resonance for very light neutralino
masses $\mchi \sim 50~\gev$. The $\tb$ dependence is also noteworthy.
The capture rate in the Earth is determined primarily by the
spin-independent elastic scattering cross section for $\chi q \to \chi
q$ through $s$-channel squarks and $t$-channel Higgs boson exchange.
All amplitudes require chirality flips, either through Higgs
interactions, squark mass insertions, or quark mass insertions. The
first two are proportional to $\tb$ and therefore dominate for
moderate and large $\tb$, leading to $C \sim \tan^2\beta$. In
contrast, for the Sun, the $\tb$ dependence is minimal; the dominant
contribution is from axial-vector scattering off Hydrogen, which is
largely independent of $\tb$.

\begin{figure}[!t]
\postscript{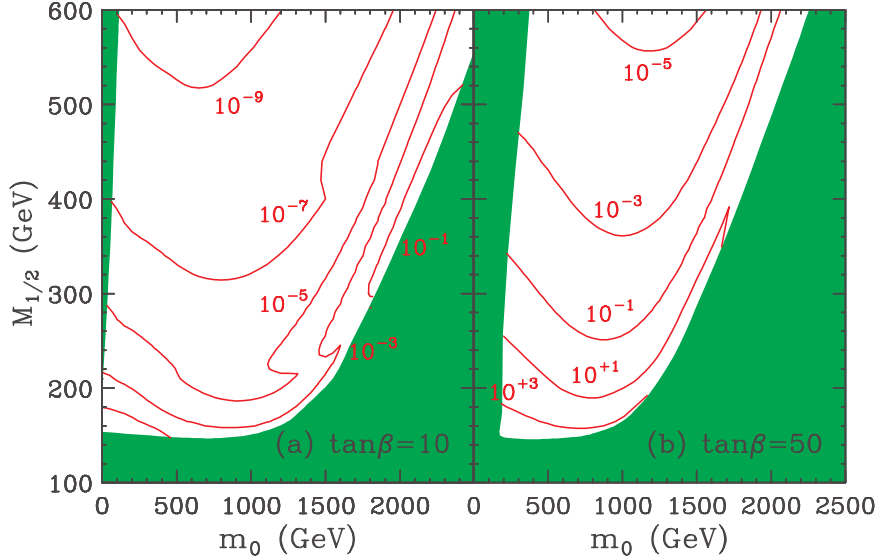}{0.70}
\caption{Muon flux from the Earth in $\km^{-2}~\yr^{-1}$
(for $\bar{v} = 270~\km/\s$ and $\rho_0 = 0.3~\gev/\cm^3$).}
\label{fig:fluxE}
\end{figure}

\begin{figure}[h]
\postscript{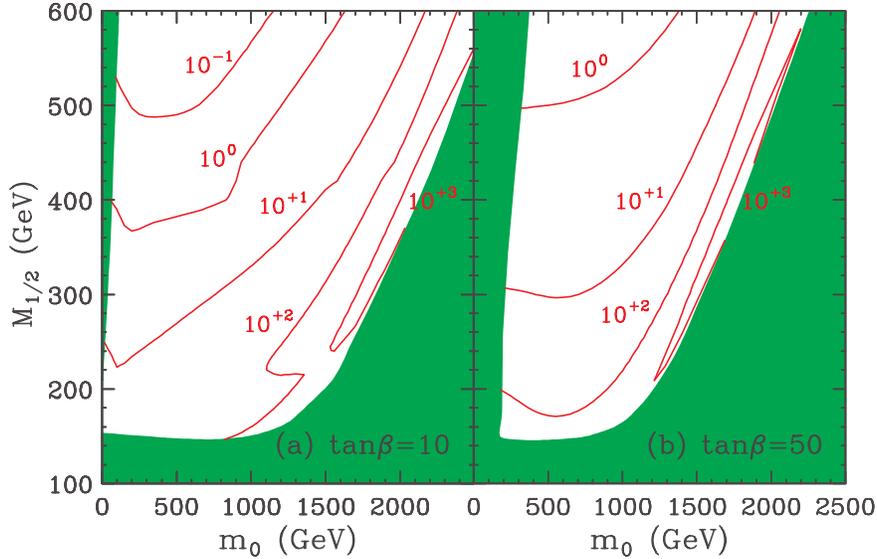}{0.70}
\caption{As in Fig.~\ref{fig:fluxE}, but for the Sun.}
\label{fig:fluxS}
\end{figure}

Muon flux rates from the Earth and Sun are presented in
Figs.~\ref{fig:fluxE} and \ref{fig:fluxS}.  Consistent with previous
studies (see, e.g. \cite{Corsetti:2000ma}),
we find that the flux rate is indeed small in regions of
parameter space with $m_0 < 1~\tev$ and Bino-like LSPs.  However, for
$m_0 > 1~\tev$, in the region where $\mchi > m_W$ and the dark matter
is a gaugino-Higgsino mixture, the fluxes are greatly enhanced.  Here,
annihilation to gauge bosons is unsuppressed, resulting in a hard
neutrino spectrum and large muon fluxes.  In this region, the rates
from the Sun are large for all values of $\tb$. For the Earth, we see
that, despite the close proximity of the Earth's center, the muon
fluxes are typically suppressed by several orders of magnitude
relative to those from the Sun.  However, reasonably large rates are
possible even for the Earth for large $\tb$, and particularly for very
light neutralinos, where the capture rate is enhanced by the iron
resonance, as discussed above.

The theoretical predictions of Figs.~\ref{fig:fluxE} and
\ref{fig:fluxS} can be compared with the experimental sensitivities of
ongoing and near future detectors~\cite{ICRC99}.  These experiments,
along with their more salient characteristics and flux limits (where
available), are listed in Table~\ref{table:NT}.  The flux limits
depend on the expected angular dispersion in the signal.  This
dispersion has two possible origins.  One is the source: although
neutralinos from the Sun are essentially a point source, in the Earth,
98\% of neutralino annihilations occur within a cone of half-angle
$8.6^{\circ} \sqrt{50~\gev/\mchi}$~\cite{Gould:1987ir,Gould:1988ww}.
The second is the angle $\theta_{\text{rms}} \approx 13^{\circ}
\sqrt{25~\gev / E_{\nu}}$ between the neutrino and its daughter muon.
As $E_{\nu} \alt \mchi/2$, $\theta_{\text{rms}}$ is typically the
dominant effect.  The flux limits listed are for half-cone sizes of
$15^{\circ}$, corresponding roughly to $\mchi \sim 50~\gev$.  Heavier
neutralinos will produce more collimated muons, allowing smaller cone
sizes with reduced backgrounds.  The improved limits for smaller cone
sizes may be found in the references.

Comparing Figs.~\ref{fig:fluxE} and \ref{fig:fluxS} with
Table~\ref{table:NT}, we find that present limits do not significantly
constrain the minimal supergravity parameter space.  However, given
that the effective area of neutrino telescope experiments is expected
to increase by 10 to 100 in the next few years, muon fluxes of order
10--100 $\km^{-2}~\yr^{-1}$ may be within reach.  Such sensitivities
are typically not sufficient to discover Bino-like LSPs, unless they
are light and $\tb$ is large.  But they have an excellent opportunity
to detect dark matter in the mixed gaugino-Higgsino dark matter
scenarios, which, as we have emphasized above, are preferred by low
energy particle physics constraints.


Muon energy thresholds, listed in Table~\ref{table:NT}, are not
included in Figs.~\ref{fig:fluxE} and \ref{fig:fluxS}. Since the muon
detection rate is dominated by high energy muons as noted above, the
threshold energy is typically not important, especially in the regions
where a detectable signal is expected.  This is not the case for all
detectors, however.  For example, since muons lose $0.26~\gev$ per
meter in water and ice, neutrino telescopes requiring track lengths of
$\sim 100$ m will have thresholds of order $\sim 25~\gev$. The
dependence on threshold energy has been studied in
Refs.~\cite{Bergstrom:1997tp,Bergstrom:1998xh}, where it was found
that for threshold energies of $E_{\mu}^{\rm thr} \sim
\frac{1}{4}\mchi$ to $\frac{1}{6}\mchi$, the loss of signal is
substantial.  Low threshold energies in neutrino telescopes are
clearly very important for dark matter detection.  This conclusion is
further strengthened by considerations of naturalness, which favor low
neutralino masses.

\begin{table}[tb]
\caption{Current and planned neutrino experiments.  We list also each
experiment's (expected) start date, physical dimensions (or
approximate effective area), muon threshold energy $E_{\mu}^{\rm thr}$
in GeV, and 90\% CL flux limits for the Earth $\Phi_{\mu}^{\oplus}$
and Sun $\Phi_{\mu}^{\odot}$ in $\km^{-2}~\yr^{-1}$ for half-cone
angle $\theta \approx 15^{\circ}$ when available.
\label{table:NT}
}
\begin{tabular}{lllrrrr}
 Experiment \rule[-2.8mm]{0mm}{7mm}
 & \hspace*{3mm} Type
  & Date
   & Dimensions\hspace*{2mm}
    & $E_{\mu}^{\rm thr}$
     & $\Phi_{\mu}^{\oplus}$\hspace*{2mm}
      & $\Phi_{\mu}^{\odot}$\hspace*{2mm}  \\  \hline
 Baksan~\cite{Baksan}
 & Underground
  & 1978
   & $17 \times 17 \times 11 \text{ m}^3$
    & 1  
     & $6.6 \times 10^3$ 
      & $7.6 \times 10^3$ \\  
 Kamiokande~\cite{Kamiokande}
 & Underground
  & 1983 
   & $\sim 150 \text{ m}^2$
    & 3
     & $10 \times 10^3$ 
      & $17 \times 10^3$ \\
 MACRO~\cite{MACRO} 
 & Underground
  & 1989
   & $12 \times 77 \times 9 \text{ m}^3$
    & $1.5^{\, \ast}$\hspace*{-2.25mm}  
     & $3.2 \times 10^3$ 
      & $6.5 \times 10^3$ \\  
 Super-Kamiokande~\cite{Okada:2000ve} 
 & Underground
  & 1996
   & $\sim 1200\text{ m}^2$
    & 1.6  
      & $1.9 \times 10^3$ 
       & $5.0 \times 10^3$  \\  
 Baikal NT-96~\cite{Baikal} 
 & Underwater
  & 1996
   & $\sim 1000\text{ m}^2$
    & 10  
      & $15 \times 10^3$ 
       &              \\  
 AMANDA B-10~\cite{AMANDAB10}
 & Under-ice
  & 1997 
   & $\sim 1000\text{ m}^{2\, \dag}$\hspace*{-1.73mm}
    & $\sim 25$
     & $44 \times 10^{3\, \dag}$\hspace*{-1.73mm}
      &     \\
 Baikal NT-200~\cite{Baikal} 
 & Underwater
  & 1998
   & $\sim 2000\text{ m}^2$
    & $\sim 10$  
     &  \\  
 AMANDA II~\cite{AMANDAII}
 & Under-ice
  & 2000
   & $\sim 3 \times 10^4\text{ m}^2$
    & $\sim 50$
     & 
      &     \\
 NESTOR$^{\S}$~\cite{NESTOR}
 & Underwater
  & 2000
   & $\sim 10^4\text{ m}^{2\, \ddag}$\hspace*{-1.73mm}
    & few
     &
      &     \\
 ANTARES~\cite{ANTARES}
 & Underwater
  & 2003
   &  $\sim 2 \times 10^4\text{ m}^{2\, \ddag}$\hspace*{-1.73mm}
    & $\sim 5$--10
     &
      &     \\    
 IceCube~\cite{AMANDAII}
 & Under-ice
  & 2003-8
   & $\sim 10^6\text{ m}^2$
    & 
     & 
      &    
\end{tabular}
$^{\ast}$ 2 GeV for Sun. \quad
$^\dag$ Hard spectrum, $\mchi = 100~\gev$. \quad
$^{\S}$ One tower. \quad
$^\ddag$ $E_{\mu} \sim 100~\gev$.
\end{table}

\section{Photons}
\label{sec:photons}

High-energy photons provide a unique signal of dark matter
annihilation.  They point back to their source, and their energy
distribution is directly measurable, at least in principle.  For these
reasons, given sufficient angular and energy resolution in gamma ray
detectors, a variety of signals may be considered.

The photon signal may arise from the galactic center
\cite{Urban:1992ej,Berezinsky:1992mx,Berezinsky:1994wv}, the galactic
halo~\cite{Bergstrom:1998zs,Bergstrom:1999jj} , or even from
extra-galactic sources~\cite{Baltz:2000ra}.  We will consider the
galactic center, where large enhancements in dark matter density are
possible~\cite{Buckley,Gondolo:1999ef}.  In contrast to the neutrino
signal considered in Sec.~\ref{sec:neutrinos}, the photon flux is
highly sensitive to halo model parameters.  Fortunately, the problem
may be separated into two parts: one containing all halo model
dependence, and the other all particle physics uncertainties.  Given
the predicted photon fluxes for a reference halo model, the
predictions for all other halo models are then easily determined.

The photon energy distribution receives two types of contributions:
line and continuum.  The former results from the loop-mediated
processes $\chi \chi \to \gamma
\gamma$~\cite{Bergstrom:1997fh,Bern:1997ng} and $\chi \chi \to \gamma
Z$~\cite{Ullio:1998ke}.  Because dark matter in the halo is extremely
non-relativistic, photons from these processes have an energy width of
only $\Delta E_{\gamma}/E_{\gamma} \sim 10^{-3}$ and are effectively
mono-energetic.  While this signal would be the most spectacular of
all possible indirect signals, its rates are, of course,
suppressed~\cite{Berezinsky:1992sn}.  In a model-independent survey,
Bergstr\"om, Ullio, and Buckley~\cite{Buckley} have found that the
photon line may be observable for neutralinos with a large Higgsino
component, assuming a cuspy halo profile, such as that of Navarro,
Frenk, and White~\cite{Navarro:1996iw}, and telescopes with small
angular acceptances $\sim 10^{-5}~\sr$.

On the other hand, photons may also be produced in the cascade decays
of other primary annihilation products. In contrast to the line
signal, cascade decays produce a large flux of photons with a
continuum of energies.  This signal is far less distinctive and will
almost certainly require additional confirmation to unambiguously
distinguish it from background or other exotic sources.  Nevertheless,
we will focus here on the continuum signal, as it will provide the
first hint of dark matter from gamma ray astronomy.

The differential photon flux along a direction that forms an angle
$\psi$ with respect to the direction of the galactic center is
\begin{equation}
\frac{d \Phi_{\gamma}}{d\Omega dE} = 
\sum_i \frac{dN_{\gamma}^i}{dE} 
\sigma_i v \frac{1}{4\pi \mchi^2} \int_{\psi} \rho^2 dl \ ,
\label{photon}
\end{equation}
where the sum is over all annihilation channels and $\rho$ is the
neutralino mass density.  All of the halo model dependence is isolated
in the integral, which, following Ref.~\cite{Buckley}, we write in the
dimensionless form
\begin{equation}
J(\psi) = \frac{1}{8.5 \text{ kpc}} \left( \frac{1}{0.3~\gev/\cm^3}
\right) ^2 \int_{\psi} \rho^2 dl \ .
\end{equation}
The integral is along the line of sight.  Assuming a spherical halo,
the mass density is given by $\rho = \rho(r)$, where $r^2 = l^2 +
R_0^2 - 2 l R_0 \cos \psi$, and $R_0 \approx 8.5$ kpc is the solar
distance to the galactic center.

The photon flux is, of course, maximized for $\psi = 0$, but it must
be averaged over the field of view.  The result is
\begin{equation}
\Phi_{\gamma} (\ethr)=  5.6 \times 10^{-10}~\cm^{-2}~\s^{-1} 
\times \sum_i \int_{\ethr}^{\mchi} \! \! dE \frac{dN_{\gamma}^i}{dE} 
\left( \frac{\sigma_i v}{\text{pb}} \right)
\left( \frac{100~\gev}{\mchi} \right)^2 
\bar{J}(\Delta \Omega) \, \Delta \Omega \ ,
\end{equation}
where
\begin{equation}
\bar{J}(\Delta \Omega) \equiv \frac{1}{\Delta \Omega} \int_{\Delta
\Omega} J(\psi)\, d\Omega \ ,
\end{equation}
and $\Delta \Omega$ is the solid angle of the field of view centered
on $\psi=0$.  $\ethr$ is the lower threshold energy; detectors also
have upper cutoffs, but these are typically irrelevant, as the energy
distribution falls steeply with energy.  $\bar{J}$ has been studied
for a variety of halo models in Ref.~\cite{Buckley}.  For a typical
atmospheric Cherenkov telescope (ACT) acceptance of $\Delta \Omega =
10^{-3}~\sr$, the modified isothermal profile described by $\rho(r)
\propto \left[ 1 + (r/a)^2 \right]^{-1}$ yields $3 \alt \bar{J} \alt
10^3$.  On the other hand, cuspy halos lead to values of $\bar{J}$ as
large as $10^5$.  (Such singular profiles have recently been argued to
be incompatible with neutralino dark matter, however, based on radio
emission from neutralino annihilation near the black hole at the
galactic center~\cite{Gondolo:2000pn}.)  We will choose a moderate
reference value $\bar{J}(10^{-3}) = 500$, which is within the allowed
ranges of both the modified isothermal and cuspy halos.  The
factorizability of the photon flux implies that our results can be
scaled to all other halo models easily.

\begin{figure}[!tp]
\postscript{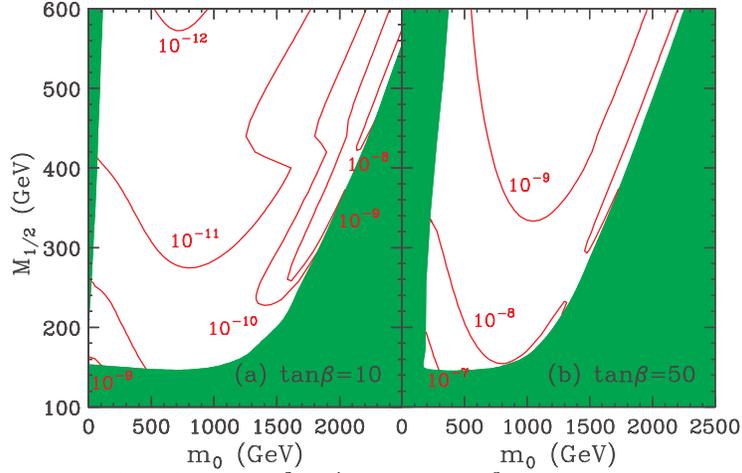}{0.59}
\caption{Photon flux $\Phi_{\gamma} (\ethr)$ in $\cm^{-2}~\s^{-1}$
from a $10^{-3}~\sr$ cone centered on the galactic center for a
threshold energy of $\ethr = 1~\gev$. We assume halo model parameter
$\bar{J} = 500$.  Results for other halo models may be obtained by
scaling to the appropriate $\bar{J}$ (see text).}
\label{fig:photon1}
\end{figure}
\begin{figure}[!hp]
\postscript{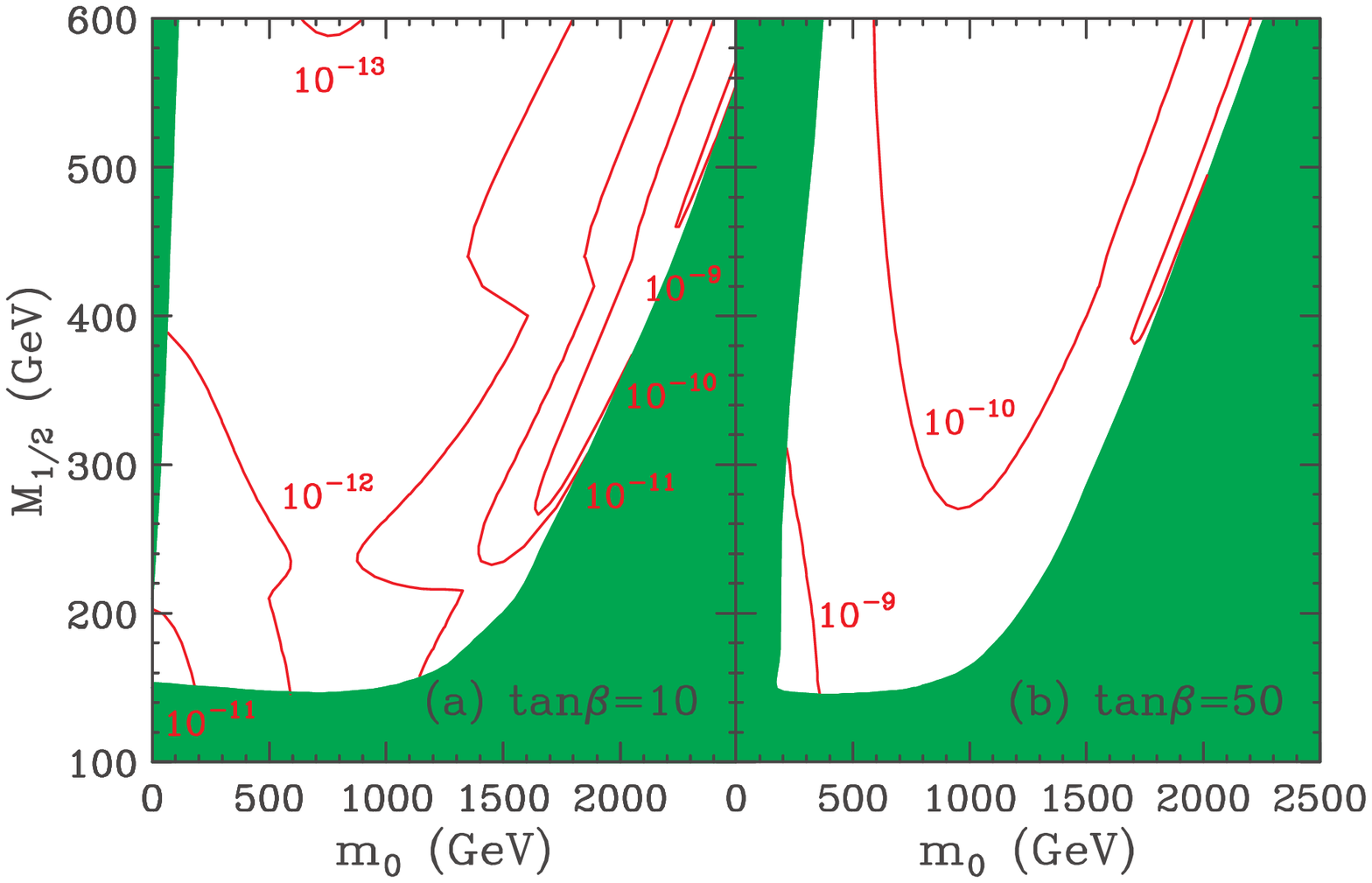}{0.59}
\caption{As in Fig.~\ref{fig:photon1}, but for the photon energy
threshold $\ethr = 10~\gev$. }
\label{fig:photon2}
\end{figure}
\begin{figure}[!hp]
\postscript{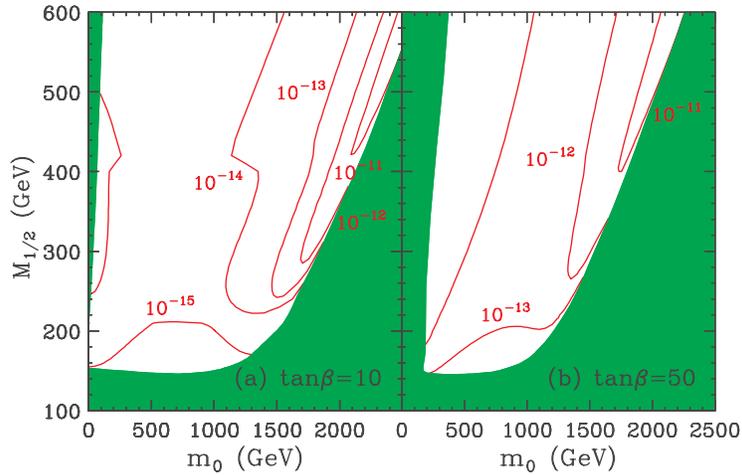}{0.59}
\caption{As in Fig.~\ref{fig:photon1}, but for the photon energy
threshold $\ethr = 50~\gev$. }
\label{fig:photon3}
\end{figure}
The particle physics model dependence enters through all the other
factors of Eq.~(\ref{photon}). The energy integral is roughly $\int
dE\, dN_{\gamma}^i / dE \sim 0.5$ for all $i$, but the energy
distribution depends significantly on the annihilation channel.  The
differential gamma ray multiplicity has been simulated for light and
heavy neutralinos in Refs.~\cite{Bengtsson:1990xf} and \cite{Buckley},
respectively.  The spectrum for the most important annihilation
channels is described well by $d N_{\gamma} / dx = a\,
e^{-bx}/x^{1.5}$, where $x \equiv E_{\gamma} / \mchi$ and $(a,b) =
(0.73, 7.76)$ for $WW$ and $ZZ$~\cite{Buckley}, $(1.0, 10.7)$ for
$b\bar{b}$, $(1.1, 15.1)$ for $t\bar{t}$, and $(0.95, 6.5)$ for
$u\bar{u}$.  We neglect Higgs boson final states, as they never have
branching fraction greater than 7\%. For the $gg$ final state, we use
the light quark distribution.  Our crude approximation for gluons is
relevant only in isolated regions with Bino-like LSPs where, as we
will see, the signal is unobservable.  With the exception of the
irrelevant light quark distribution, the $WW$ and $ZZ$ distributions
produce the most energetic photons.

The photon flux $\Phi_{\gamma} (\ethr)$ is given in
Figs.~\ref{fig:photon1}--\ref{fig:photon3} for threshold energies of
1, 10 and 50 GeV.  The maximal rates are found in the region of
parameter space with mixed gaugino-Higgsino dark matter, and are
insensitive to $\tb$.  Here branching ratios to gauge bosons are
large, and the photon spectrum hard.  In the rest of parameter space,
$b\bar{b}$ is an important final state, and $\Phi_{\gamma} (\ethr)$ is
enhanced by $\tb$.

In the past, the high energy photon spectrum with $10~\gev \alt
E_{\gamma} \alt 300~\gev$ has been largely unexplored. Ground-based
detectors, such as the Whipple 10m telescope, have large effective
areas, but have traditionally been limited to energies above $\sim
300~\gev$.  Space-based detectors, such as EGRET, have been sensitive
to photon energies up to $\sim 20~\gev$, but are limited above this
energy by their small effective area.  There has therefore been an
unexplored gap at intermediate energies, which happens to overlap
substantially with the range of energies most favored by
supersymmetric dark matter.

\begin{table}[tb]
\caption{Some of the current and planned $\gamma$ ray detector
experiments with sensitivity to photon energies $10~\gev \protect\alt
E_{\gamma} \protect\alt 300~\gev$.  We list each experiment's
(proposed) start date and expected $E_{\gamma}$ coverage in GeV.  The
energy ranges are approximate.  For experiments constructed in stages,
the listed threshold energies will not be realized initially.  See the
references for details.
\label{table:photons}
}
\begin{tabular}{lllr}
 Experiment \rule[-2.8mm]{0mm}{7mm}
  & \hspace*{2mm} Type
   & Date
    & $E_{\gamma}$ Range
\\  
\hline
 EGRET~\cite{EGRET}
  & Satellite
   & 1991-2000
    & 0.02--30
\\
 STACEE~\cite{STACEE}
  & ACT array
   & 1998
    & 20--300
\\
 CELESTE~\cite{CELESTE}
  & ACT array
   & 1998
    & 20--300
\\
 ARGO-YBJ~\cite{ARGO-YBJ}
  & Air shower
   & 2001
    & 100--2,000
\\
 MAGIC~\cite{MAGIC}
  & ACT
   & 2001
    & 10--1000
\\
 AGILE~\cite{AGILE}
  & Satellite
   & 2002
    & 0.03--50
\\
 HESS~\cite{HESS}
  & ACT array
   & 2002
    & 40--5000
\\
 AMS/$\gamma$~\cite{AMSgamma}
  & Space station
   & 2003
    & 0.3--100
\\
 CANGAROO III~\cite{CANGAROOIII}
  & ACT array
   & 2004
    & 30--50,000
\\
 VERITAS~\cite{VERITAS}
  & ACT array
   & 2005
    & 50--50,000
\\
 GLAST~\cite{GLAST}
  & Satellite
   & 2005
    & 0.1--300
\end{tabular}
\end{table}
The experimental situation is changing rapidly, however.  Currently,
two heliostat arrays, STACEE and CELESTE, are running with sensitivity
in the range $20~\gev \alt E_{\gamma} \alt 300~\gev$, and many more
experiments with greatly improved sensitivity are expected in the next
few years.  Upcoming experiments with sensitivity to $\gamma$ rays
with $10~\gev \alt E_{\gamma} \alt 300~\gev$ are listed in
Table~\ref{table:photons}.

\begin{figure}[t]
\postscript{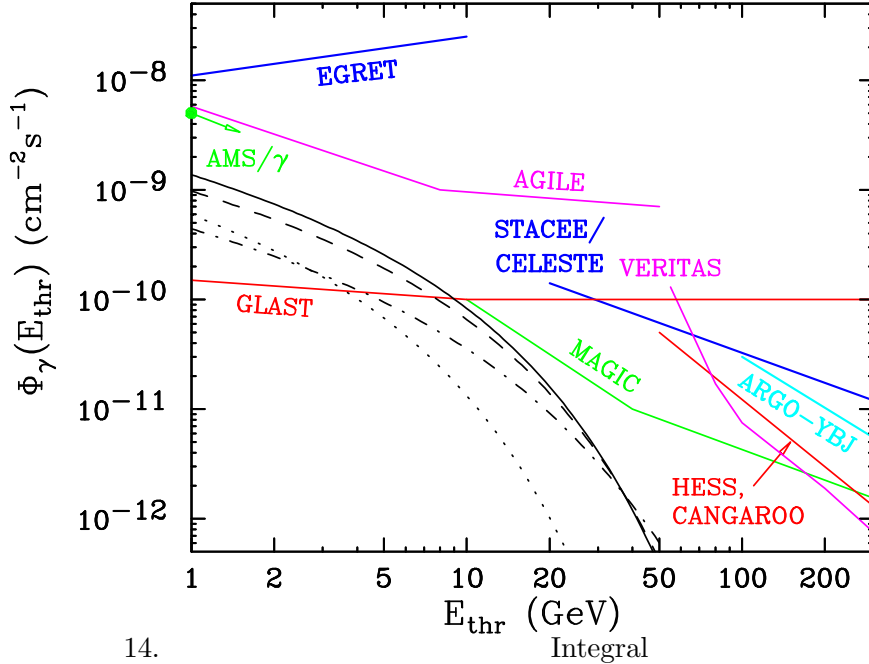}{0.70}
\caption{Integral photon fluxes $\Phi_{\gamma}(\ethr)$ as a function of
the threshold energy $\ethr$ for $A_0=0$, $\mu>0$, $m_t = 174~\gev$, and
halo parameter $\bar{J} = 500$.  The four models have relic density
$\Omegachi h^2 \approx 0.15$, and are specified by
$(\tan\beta, m_0, M_{1/2}, m_{\chi}, R_{\chi}) = (10, 100, 170, 61, 0.93)$ (dotted),
$(10, 1600, 270, 97, 0.77)$ (dashed), $(10, 2100, 500, 202, 0.88)$
(dot-dashed), and $(50, 1000, 300, 120, 0.96)$ (solid), where all
masses are in GeV. Point source flux sensitivity estimates for several
gamma ray detectors are also shown. (Care should be taken in comparing
these sensitivities to the predicted fluxes --- see text.)  }
\label{fig:photon_explims}
\end{figure}

An important figure of merit for the detection of $\gamma$ rays from
the galactic center is the point source flux sensitivity.  A
compilation of previous estimates of flux sensitivities is given in
Fig.~\ref{fig:photon_explims} for EGRET, STACEE, CELESTE, ARGO-YBJ,
MAGIC, AGILE, HESS~\cite{HESS-2}, AMS/$\gamma$~\cite{AMSgamma}, 
VERITAS~\cite{VERITAS}, and GLAST~\cite{GLAST}.
The flux sensitivities for the first six experiments are from
Ref.~\cite{AGILE}, and those for the remaining experiments are from
the references listed.  The sensitivity of MAGIC assumes the
availability of high quantum efficiency photosensors. The sensitivity
for CANGAROO III is currently being re-evaluated~\cite{CANGAROOIII-2}.
We have included it in accord with expectations that it will be
comparable to that of HESS.
The point flux sensitivities are, of course, dependent
on the source's location and energy spectrum.
They are also subject to a variety of other
experimental uncertainties and assumptions; see the references for
details. A typical estimate~\cite{VERITAS} assumes background
extrapolated from EGRET data~\cite{EGRET}, and a signal distribution
$dN_{\gamma}/dE \propto E^{-2.5}$.  Detector efficiencies and cuts are
included, and a 5$\sigma$ signal with at least 10 photons is required.
50 hours of observation is assumed for telescopes, and one year of an
all sky survey for the space-based detectors.  The arrow for
AMS/$\gamma$ in Fig.~\ref{fig:photon_explims} indicates that a
published estimate exists only for $\ethr = 1~\gev$, but flux
sensitivity at some level can be expected out to the detector limit of
100 GeV.

The experimental sensitivities presented clearly cannot be interpreted
as future dark matter discovery contours.  The neutralino signal has a
different energy spectrum than assumed, and the background in the    
direction of the galactic center is larger, due to the diffuse
$\gamma$ ray emission from the galactic disk~\cite{Chardonnet:1995db},
which enhances $\sqrt{B}$ by a factor of $\sim 5$
\cite{egret:galactic}. (This last fact implies that for
some halo profiles, it may be advantageous to center the field of view
away from the galactic center. This optimization may significantly
reduce potential losses in signal significance.)  In addition, there
are many ambiguities in background calibration, and, as noted above,
the continuum signal is not sufficiently distinct for a simple excess
to identify the source as dark matter annihilation.  However, the flux
sensitivities of Fig.~\ref{fig:photon_explims} do clearly portend
substantial progress in the next few years, and can serve as rough
indications of what signal levels will be detectable.

The expected fluxes for four supersymmetry models and $\bar{J} = 500$
are also shown in Fig.~\ref{fig:photon_explims}.  Although there is a
large uncertainty from halo model dependence, it is clear that
detectable signals are possible.  At present, EGRET data is not overly
constraining, although halo models with large $\bar{J} \sim 5000$ are
within EGRET sensitivity and may even explain a flattening of the
spectrum.  In the future, AMS/$\gamma$ and AGILE will improve this
sensitivity, and MAGIC may see excesses for
$\bar{J} \sim 500$.  Finally, GLAST will provide the greatest
sensitivity of all, probing halo models with $\bar{J}$ as low as
$\bar{J} \sim 50$.

If a significant excess is found in future experiments, its dark
matter origin can be tested in a variety of ways.  Confirmation from
other searches for dark matter or supersymmetry would be the most
satisfying possibility.  As we will see in Sec.~\ref{sec:comp}, the
neutrino and positron signals probe similar models, so a coincidence
of various signals is a distinct possibility. However, it has also
been suggested that an angular distribution of photons that does not
follow the galactic disk and bulge may be a powerful
diagnostic~\cite{Buckley}.  Also, as the dark matter signal has a
shape differing from the background, detailed likelihood fits to the
photon energy distribution may also be a useful tool, although far
beyond the scope of this work.  It seems clear, in any case, that for
reasonable halo models and supersymmetry parameters, meaningful
$\gamma$ ray signals in the next few years are possible, particularly
with gaugino-Higgsino dark matter.

\section{Positrons}
\label{sec:positrons}

An excess of cosmic anti-particles and anti-matter from dark matter
annihilation may be detected in space-based or balloon-borne
experiments.  The positron signal is perhaps the most
promising~\cite{Tylka:1989xj,Turner:1990kg,Kamionkowski:1991ty,%
Baltz:1999xv,Moskalenko:1999sb}.  In the past, soft anti-protons with
energies $\sim 100$ MeV have also been
considered~\cite{Stecker:1988fx,Chardonnet:1996ca}.  However, recent
work finds larger background than previously expected, complicating
the identification of an anti-proton
signal~\cite{Bergstrom:1999jc,Bieber:1999dn}.  Anti-deuterium has also
been suggested as a possibility~\cite{Donato:1999gy}.

The positron background is most likely to be composed of secondaries
produced in the interactions of cosmic ray nuclei with interstellar
gas, and is expected to fall as $\sim E_{e^+}^{-3.1}$.  At energies
below 10 GeV, however, this background is subject to large
uncertainties from the effects of the solar
wind~\cite{Baltz:1999xv,Moskalenko:1999sb}.  The soft positron
spectrum also varies depending on the orbit path of the experiment.
At high energies, these effects are negligible.  In addition,
positrons lose energy through a variety of processes, and so hard ones
must typically be produced within a few
kpc~\cite{Baltz:1999xv,Moskalenko:1999sb}.  For this reason, the hard
spectrum is relatively insensitive to variations in the halo profile
near the galactic center.  The dark matter signal is therefore most
promising at high energies, where the background is relatively small
and well understood.

The differential positron flux is~\cite{Moskalenko:1999sb}
\begin{equation}
\frac{d\Phi_{e^+}}{d\Omega dE} = \frac{\rho_0^2}{\mchi^2}
\sum_i \sigma_i v B_{e^+}^i 
\int dE_0\, f_i(E_0)\, G(E_0, E) \ ,
\end{equation}
where $\rho_0$ is the local neutralino mass density, the sum is over
all annihilation channels, and $B_{e^+}^i$ is the branching fraction
to positrons in channel $i$. The source function $f(E_0)$ gives the
initial positron energy distribution from neutralino annihilation.
$G(E_0, E)$ is the Green's function describing positron propagation in
the galaxy~\cite{Protheroe:1982pp} and contains all the halo model
dependence.

For the reasons mentioned above, processes yielding hard positrons are
by far the most important for dark matter discovery.  The `positron
line' signal from $e^+ e^-$ is helicity-suppressed.  It may be
enhanced, for example, in the case of Bino-like LSPs if selectrons are
much lighter than all other scalars, but this possibility is highly
unmotivated. To an excellent approximation, then, hard positrons arise
from $\chi \chi \to WW, ZZ$, followed by the direct decay of gauge
bosons to positrons. Assuming unpolarized gauge bosons, $f$ is the
familiar flat distribution with endpoints determined by the gauge
boson and neutralino masses. The Green's function $G$ has been modeled
by Moskalenko and Strong in Ref.~\cite{Moskalenko:1999sb} in a
framework that consistently reproduces a wide range of observational
data from anti-protons, nuclei, electrons, positrons, and photons.

Combining all of these results, the differential positron flux may be
written as
\begin{eqnarray}
E^2 \frac{d\Phi_{e^+}}{d\Omega dE} &=& 
0.027~\cm^{-2}~\s^{-1}~\sr^{-1}~\gev \nonumber \\
&& \times \left( \frac{\rho_0}{0.3~\gev/\cm^3} \right)^2 
\left( \frac{100~\gev}{\mchi} \right)^2 
\sum_i \frac{\sigma_i v}{\text{pb} \cdot \beta_i} B_{e^+}^i
\int_{z_-^i}^{z_+^i} dz\, g(z, E/\mchi) \ ,
\label{diffpositron}
\end{eqnarray}
where 
\begin{eqnarray}
\beta_{WW, ZZ} &=& \left(1- m_{W,Z}^2/\mchi^2\right)^{1/2} \ , \\
z_{\pm}^i &=& \left(1\pm \beta_i \right)/2 \ , \\
B_{e^+}^{WW} &=& B(W^+ \to e^+ \nu) = 0.11 \ , \\
B_{e^+}^{ZZ} &=& 2B(Z\to e^+ e^-) = 0.067 \ ,
\end{eqnarray}
and the reduced Green's function is
\begin{equation}
g(z, E/\mchi) \equiv
 10^{a \log_{10}^2 E+ b\log_{10} E +c}\ \theta(z -E/\mchi)
+10^{w \log_{10}^2 E+ x\log_{10} E +y}\ \theta(E/\mchi - z) \ ,
\end{equation}
where $E$ is in GeV and the ($z$-dependent) coefficients $a,b,c$ and
$w,x,y$ are tabulated in Ref.~\cite{Moskalenko:1999sb} for different
halo profiles.  As mentioned above, at high energies, these
coefficients are fairly independent of the halo model, as high energy
positrons originate in our solar neighborhood, where all profiles give
similar densities. We adopt coefficients corresponding to the modified
isothermal distribution with halo size 4 kpc. For large $\mchi$, the
integral of Eq.~(\ref{diffpositron}) is insensitive to $\mchi$, and so
the differential positron flux scales as $\sim 1/\mchi^4$.
Neutralinos with mass not far above $m_W$ are therefore most easily
detected.

\begin{figure}[t]
\postscript{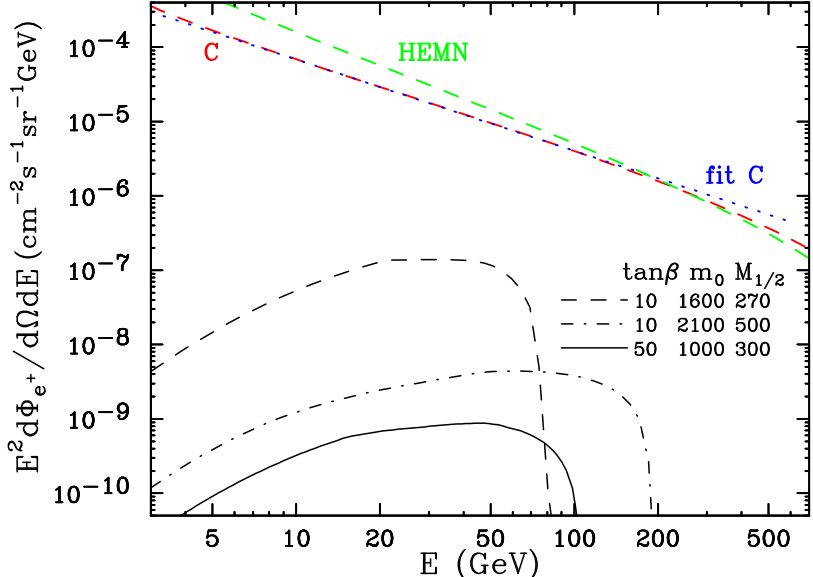}{0.65}
\caption{The differential positron flux for three of the four sample
points in Fig.~\ref{fig:photon_explims}.  The curves labeled C and
HEMN are background models from Ref.~\protect\cite{Moskalenko:1999sb};
the dotted curve is our fit to C.}
\label{fig:dFdEPos}
\end{figure}

In Fig.~\ref{fig:dFdEPos}, we show three sample spectra for
supersymmetry models yielding relic abundances of $\Omegachi h^2
\approx 0.15$.  Two background spectra from
Ref.~\cite{Moskalenko:1999sb} are also shown.  The signal rates are
significantly suppressed relative to those of
Refs.~\cite{Kamionkowski:1991ty,Moskalenko:1999sb}, where the dark
matter was assumed to be Higgsino-like.  Higgsino-like dark matter is
highly disfavored, however, as, unless it is unnaturally heavy, it
annihilates too strongly to leave interesting relic abundances. As is
evident from Fig.~\ref{fig:gfrac}, in the allowed minimal supergravity
parameter space the LSP is far from pure Higgsino-like, particularly
in the region with preferred relic density.

\begin{figure}[ht]
\postscript{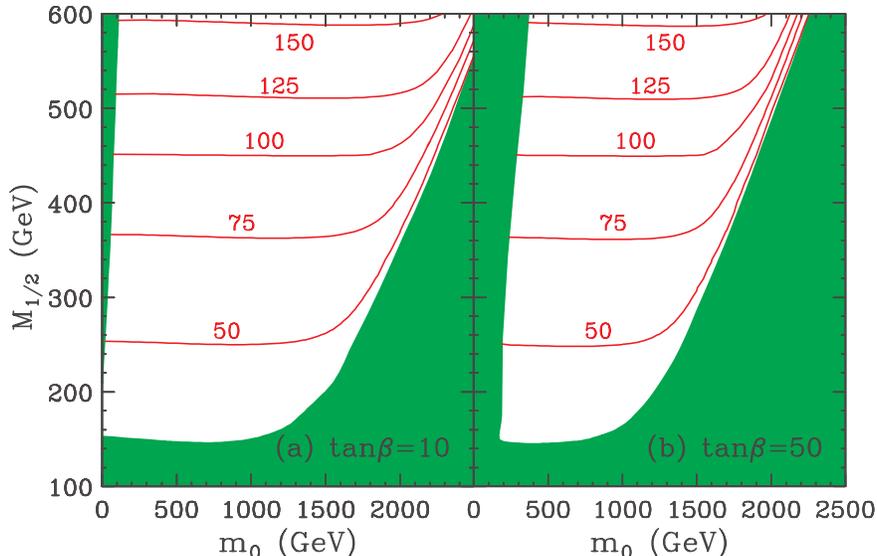}{0.70}
\caption{The optimal positron energy $\eopt$ in GeV at which the
signal to background ratio $S/B$ is maximized. }
\label{fig:emax_Pos}
\end{figure}
\begin{figure}[ht]
\postscript{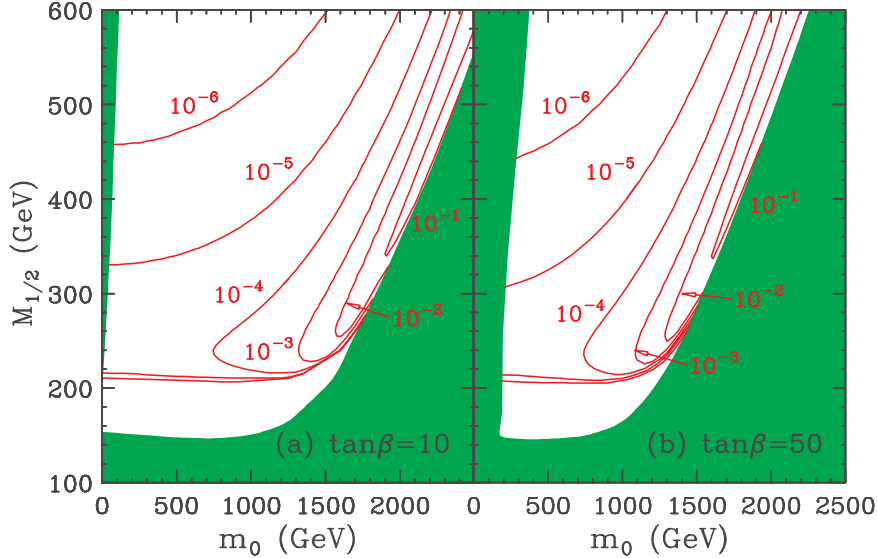}{0.70}
\caption{The positron signal to background ratio $S/B$ at $\eopt$. }
\label{fig:sob_Pos}
\end{figure}

As positrons result from two-body decay, we expect the signal, and the
signal to background ratio $S/B$, to be peaked near $\mchi/2$.  This
is evident in the three examples give in Fig.~\ref{fig:dFdEPos}.  In
Fig.~\ref{fig:emax_Pos}, we plot the optimal energy $\eopt$ at which
the signal to background ratio is maximized.  Our fit to background C
is $E^2 d\Phi_{e^+}/d\Omega dE = 1.16\times 10^{-3} E^{-1.23}$, where
$E$ is in GeV.  Comparing with Fig.~\ref{fig:mlsp}, we see that
$\eopt$ is indeed approximately $\mchi/2$ throughout parameter space.
In Fig.~\ref{fig:sob_Pos}, we plot $S/B$ at $\eopt$.  $S/B$ is
substantial only in the gaugino-Higgsino region with $\mchi > m_W$.

Figs.~\ref{fig:emax_Pos} and \ref{fig:sob_Pos} imply that the best
experimental hope for indirect detection of dark matter through
positrons is in experiments sensitive to positron energies above $\sim
50~\gev$.  In the next two to three years, both PAMELA, a satellite
detector, and AMS-02, an experiment to be placed on the International
Space Station, will satisfy this requirement.  These experiments and
other recently completed experiments are listed in
Table~\ref{table:positrons}.

\begin{table}[tb]
\caption{Recent and planned $e^+$ detector experiments. We list each
experiment's (expected) start date, duration, geometrical acceptance
in $\cm^2~\sr$, maximal $E_{e^+}$ sensitivity in GeV, and (expected)
total number of $e^+$ detected per GeV at $E_{e^+} = 50$ and 100 GeV.
\label{table:positrons}
}
\begin{tabular}{lllrrrrr}
 Experiment \rule[-2.8mm]{0mm}{7mm}
  & Type
   & Date
    & Duration
    & Acceptance
     & $E_{e^+}^{\text{max}}$
      & $\frac{dN}{dE}(50)$
       & $\frac{dN}{dE}(100)$ \\  \hline
 HEAT94/95~\cite{HEAT}
  & Balloon
   & 1994/95
    & 29/26 hr
    & 495
     & 50
      & ---
       & ---          \\
 CAPRICE94/98~\cite{CAPRICE}
  & Balloon
   & 1994/98
    & 18/21 hr
    & 163
     & 10/30
      & ---
       & ---          \\
 PAMELA~\cite{PAMELA}
  & Satellite
   & 2002-5
    & 3 yr
    & 20
     & 200
      & 7
       & 0.7         \\
 AMS-02~\cite{AMS-02}
  & Space station
   & 2003-6
    & 3 yr
    & 6500
     & 1000
      & 2300
       & 250
\end{tabular}
\end{table}
The expected number of positrons per GeV are listed in
Table~\ref{table:positrons} at positron energies of 50 and 100 GeV.
After integrating over some appropriate energy bin size, we see that
the expected statistical errors are roughly $\sim 10\%$ for PAMELA,
and $\sim 1\%$ for AMS-02.  Of course, the signal will also be
degraded by systematic errors, particularly in the background
calculation.  It seems likely, however, that the characteristic
peaking of the dark matter signal over an interval of ${\cal
O}(10~\gev)$ will be distinctive.  In addition, some systematic errors
may be eliminated by considering the ratio $e^+/(e^- + e^+)$.  While
the positron signal is typically too small for Bino-like LSPs, an
excess of $\sim 1\%$ is possible for gaugino-Higgsino dark matter. The
region of detectable positron signals may be extended, however, if,
for example, the halo is clumpy, or if the local density is larger
than our reference value of $0.3~\gev/\cm^3$~\cite{Baltz:1999xv}.

\section{Comparison with Other Searches}
\label{sec:comp}

In the previous three sections, we have examined several indirect
signals of neutralino dark matter.  As emphasized in
Sec.~\ref{sec:introduction}, however, supersymmetric dark matter
cannot exist in isolation, and there are many other avenues for
probing supersymmetric models.  We now discuss several other promising
probes and their projected sensitivities, and we then compare their
reaches.

Most closely linked to indirect searches are searches for dark matter
scattering off nuclei in low-background detectors. The DAMA
collaboration has reported evidence for an annual modulation
signal~\cite{Bernabei:2000qi}, and the activity in this field will
intensify tremendously in the next few years.  (For a recent review,
see, {\em e.g.}, Ref.~\cite{Baudis:1999aa}.)  Here we will use our
previous results~\cite{FMW} to estimate the sensitivities of the
direct searches. We choose CDMS (Soudan)~\cite{Schnee:1998gf} and
CRESST~\cite{Bravin:1999fc} as examples of near-term future
experiments.  Their projected sensitivities in neutralino-proton cross
section are both of order $\sigma_{\text{proton}} \sim
10^{-8}~\text{pb}$ for $50~\gev \alt \mchi \alt 500~\gev$.  More
precisely, we parameterize their sensitivities as
\begin{equation}
\sigma_{\text{proton}}\ \simeq\
\exp\left\{a + b \left({m_\chi\over 100~\gev}\right)
 + c \left({m_\chi\over 100~\gev}\right)^2 \right\}~\text{pb} \ ,
\label{direct}
\end{equation}
with $(a, b, c) = (-17, -4.5, 3.1)$ for $m_\chi < 84~\gev$ and $(-19,
0.68, -0.057)$ for $m_\chi>84~\gev$.  This limit may be improved by an
order of magnitude by the recently proposed GENIUS
project~\cite{Klapdor-Kleingrothaus:1999hk}, or even by CRESST itself,
assuming three years of operation with improved background
rejection~\cite{Bravin:1999fc}.

Among high energy colliders, the Large Hadron Collider (LHC) at CERN
is the ultimate supersymmetry discovery machine and will discover at
least some superpartners in all of the regions of parameter space we
have plotted.  The LHC is scheduled to begin operation in its low
luminosity mode in 2006. Before that, however, both the LEP II and
Tevatron colliders have a chance to discover superpartners.  The most
stringent constraint from LEP II on minimal supergravity comes from
chargino searches. LEP II is now concluding its run, and by the end of
2000 will improve the current chargino mass limits by about 5 GeV. If
no signal is seen, this will marginally extend the bottom and right
excluded regions of our figures.

The Tevatron will begin operation early in 2001.  In the first two
years, Run IIa will provide an integrated luminosity of $2~\ifb$ for
each detector before a temporary shutdown for a year of detector
maintenance and upgrades. In the subsequent Run IIb, the data
acquisition rate is expected to be about $5~\ifb$ per year per
detector. Hence, by 2006 we expect $10-12~\ifb$ for each Tevatron
collaboration. The Tevatron supersymmetry reach has been extensively
studied recently~\cite{Abel:2000vs}, with the conclusion that there is
some sensitivity, but in rather limited regions of parameter space.
The most effective signal is in the clean trilepton
channel~\cite{Matchev:1999nb,Baer:2000bq,Barger:1999hp,Matchev:1999yn}
resulting from chargino-neutralino pair production, followed closely
by the jets plus $\met$ channel~\cite{Baer:1998bj} and the dileptons
plus $\tau$ jet channel~\cite{Lykken:2000kp}. The maximal reach in
chargino mass is $170~\gev$ in the Bino LSP region at very low $m_0$,
where the leptonic branching ratios of the electroweak gauginos are
enhanced by light sleptons. This degrades rapidly at higher $m_0$,
where hadronic decays are prominent.  It also requires moderate $\tb$.
At large values of $\tan\beta$, decays to $\tau$ leptons dominate the
small $m_0$ region and signatures with $\tau$ jets must be
used~\cite{Baer:1998bj,Lykken:2000kp}.

At present there are no dedicated Tevatron studies in the focus point
region.  (For an LHC study, see Ref.~\cite{Allanach:2000ii}.) There
are several important modifications to collider signals for $m_0 >
1~\tev$.  For example, the lighter chargino and neutralinos are more
degenerate, leading to softer decay products, and their branching
ratios to $b$ quarks are enhanced by their Higgsino component.  Such
issues may have a large impact on chargino and neutralino searches at
the Tevatron.  This is an important question, but currently the
Tevatron reach in the focus point region is unknown.

While the Higgs boson is not a supersymmetric particle, supersymmetry
(in its economical implementations) restricts its mass, and so Higgs
boson searches also have an important impact on supersymmetric
models. For LEP II, the ultimate exclusion limit, barring a discovery,
is expected to be $m_h > 115~\gev$.  At the Tevatron, the $3\sigma$
($5\sigma$) Higgs boson discovery reach for $10~\ifb$ is $m_h\alt 100
\ (120)$ GeV~\cite{higgsreport,Baer:1999rg,Carena:1999gk}.  The Higgs
boson mass, unlike all other quantities investigated here, is
sensitive to the $A_0$ parameter. As the Higgs boson mass limit rises,
models with non-zero $A_0$, large $\tb$, and $m_0 \agt 1~\tev$ are
increasingly favored~\cite{FMM2}.  However, for natural values of
$A_0$~\cite{FMM2}, $100~\gev < m_h < 120~\gev$ and so the Higgs boson
will be discovered at the Tevatron at $3\sigma$, but never at
$5\sigma$.

Finally, there are many opportunities for discovering supersymmetry in
low energy experiments. These include effects in hadronic and leptonic
flavor violation, CP violation, proton decay, and electric and
magnetic dipole moments.  These are discussed more completely in
Ref.~\cite{FM}.  Here we will focus on two particularly robust probes:
$B \to X_s \gamma$ and the anomalous magnetic moment of the muon.

The best current measurements of $B\rightarrow X_s\gamma$ from CLEO
and ALEPH can be combined in a weighted average of $B(B\rightarrow
X_s\gamma)_{\rm exp} =(3.14\pm 0.48)\times
10^{-4}$~\cite{Kagan:1999ym}.  These measurements will be improved at
the $B$ factories, where large samples of $B$ mesons will greatly
reduce statistical errors.  However, the uncertainty in the
theoretical prediction of the standard model, $B(B\rightarrow
X_s\gamma)_{\rm SM} =(3.29\pm 0.30)\times 10^{-4}$, is likely to
remain unchanged. By 2006, a conservative estimate is that both
theoretical and experimental uncertainties will be $\sim 0.3 \times
10^{-4}$. Combining them linearly, the $2\sigma$ limit will be
$2.1\times 10^{-4} < B(B\rightarrow X_s\gamma) < 4.5\times 10^{-4}$.

The supersymmetric contribution to the muon magnetic dipole moment
(MDM) $a_{\mu} = \frac{1}{2}(g-2)_{\mu}$ is also a robust probe, since
it involves only a few (flavor- and CP-conserving)
parameters~\cite{Moroi:1996yh}.  The world average is $a_\mu^{\rm exp}
= (116\ 592\ 05\pm 45)\times 10^{-10}$~\cite{Carey} and is consistent
with the standard model.  However, once data currently being taken is
analyzed, the Brookhaven experiment E821 is expected to reduce the
uncertainty to $\Delta a_{\mu} \sim 4 \times
10^{-10}$~\cite{GrossePerdekamp:1999up}.  At present, uncertainties in
the standard model prediction are substantial.  Assuming these can be
reduced, however, a reasonable estimate for future $2\sigma$
sensitivity is $a_\mu^{\text{SUSY}} = 8 \times 10^{-10}$.

\begin{table}[tb]
\caption{Constraints on supersymmetric models used in
Figs.~\ref{fig:reach10} and \ref{fig:reach50}.  We also list
experiments likely to reach these sensitivities before 2006.  
\label{table:comp}
}
\begin{tabular}{llll}
 Observable \rule[-2.8mm]{0mm}{7mm}
  & Type
   & Bound
    & Experiment(s)   \\  \hline
 $\tilde{\chi}^+ \tilde{\chi}^-$
  & Collider
   & $m_{\tilde{\chi}}^{\pm} > 100~\gev$
    & LEP: ALEPH, DELPHI, L3, OPAL     \\ 
 $\tilde{\chi}^{\pm} \tilde{\chi}^0$
  & Collider
   & See Refs.~\cite{Matchev:1999nb,Matchev:1999yn,Lykken:2000kp}
    & Tevatron: CDF, D0  \\ 
 $B \to X_s \gamma$
  & Low energy
   & $|\Delta B(B\rightarrow X_s\gamma)| < 1.2\times 10^{-4}$
    & BaBar, BELLE     \\
 Muon MDM
  & Low energy
   & $|a_{\mu}^{\text{SUSY}}| < 8 \times 10^{-10}$
    & Brookhaven E821  \\
 $\sigma_{\text{proton}}$
  & Direct DM
   & Equation~(\ref{direct})
    & CDMS, CRESST, GENIUS \\
 $\nu$ from Earth
  & Indirect DM
   & $\Phi_{\mu}^{\oplus} < 100~\km^{-2}~\yr^{-1}$
    & AMANDA, NESTOR, ANTARES \\
 $\nu$ from Sun
  & Indirect DM
   & $\Phi_{\mu}^{\odot} < 100~\km^{-2}~\yr^{-1}$
    & AMANDA, NESTOR, ANTARES \\
 $\gamma$ (gal. center)
  & Indirect DM
   & $\Phi_{\gamma}(1) < 1.5\times 10^{-10}~\cm^{-2}~\s^{-1}$
    & GLAST \\
 $\gamma$ (gal. center)
  & Indirect DM
   & $\Phi_{\gamma}(50) < 7\times 10^{-12}~\cm^{-2}~\s^{-1}$
    & MAGIC \\
 $e^+$ cosmic rays
  & Indirect DM
   & $(S/B)_{\text{max}} < 0.01$
    & AMS-02           
\end{tabular}
\end{table}
In Table~\ref{table:comp} we present our estimates for sensitivities
that will be achieved before the LHC begins operation.  The
experiments likely to achieve these projections are also listed.
Using these estimates, the reach in minimal supergravity parameter
space for each mode is given in Figs.~\ref{fig:reach10} and
\ref{fig:reach50}.  In reading these figures, recall that we have
assumed constant local densities in our assessment of dark matter
search reaches.  If one assumes that the local density is modulated by
the thermal relic density, the dark matter reaches outside the shaded
regions should be suitably diminished.  Within the shaded regions,
however, our analysis applies without modification.

Several striking features emerge from Figs.~\ref{fig:reach10} and
\ref{fig:reach50}.  First, we see that, within the minimal
supergravity framework, nearly all of the cosmologically preferred
models will be probed by at least one experiment.  This is strictly
true for $\tb=10$.  For $\tb=50$, some of the preferred region escapes
all probes, but this requires $\mgaugino \agt 450~\gev$ and $m_0 \agt
1.5~\tev$, and requires significant fine-tuning of the electroweak
scale. In the most natural regions, all models in which neutralinos
form a significant fraction of dark matter will yield some signal
before the LHC begins operation.

\begin{figure}[!tp]
\postscript{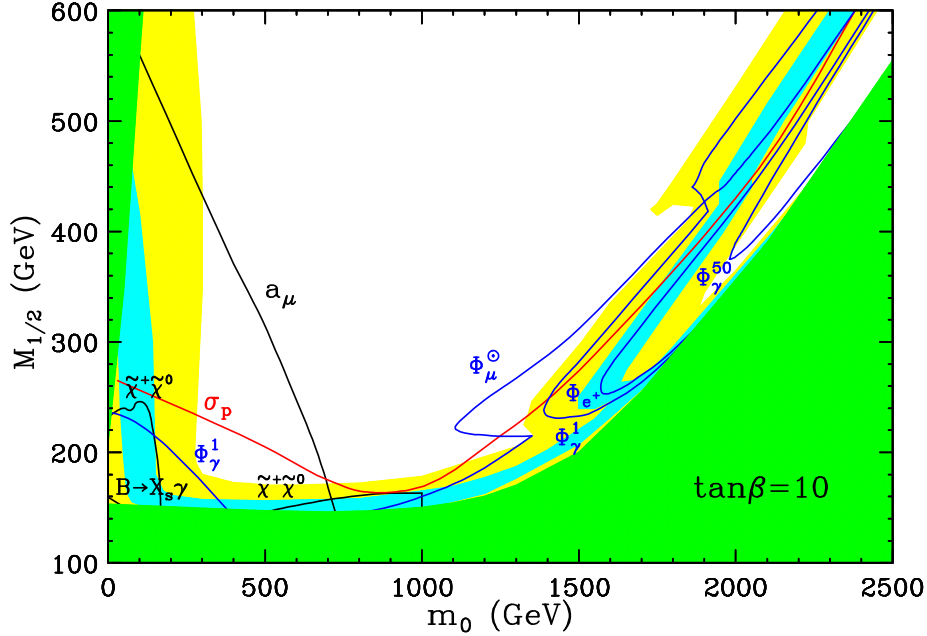}{0.74}
\caption{Estimated reaches of various high-energy collider and
low-energy precision searches (black), direct dark matter searches
(red), and indirect dark matter searches (blue) before the LHC begins
operation, for $\tb=10$. The projected sensitivities used are given in
Table~\ref{table:comp}. (The LEP chargino mass bound will marginally
extend the bottom and right excluded regions and is omitted.) The
shaded regions are as in Fig.~\ref{fig:mlsp}. The regions probed
extend the curves toward the forbidden, green regions. The dark matter
reaches are {\em not} modulated by the thermal relic density.  Bounds
from photons from the galactic center are highly halo model-dependent;
we assume a moderate halo profile parameter $\bar{J} = 500$. (See
text.)}
\label{fig:reach10}
\end{figure}
\begin{figure}[!hp]
\postscript{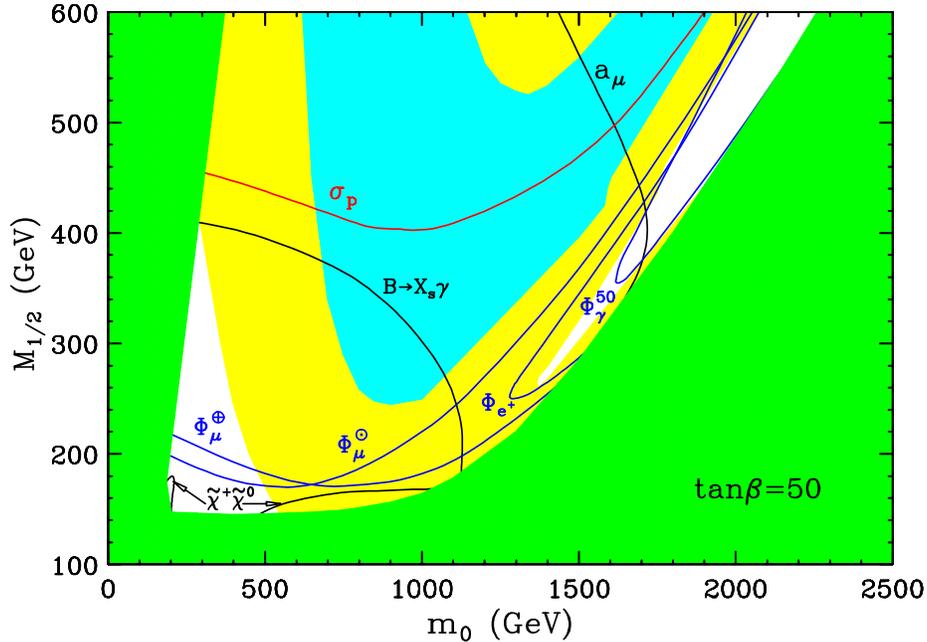}{0.74}
\caption{As in Fig.~\ref{fig:reach10}, but for $\tb=50$. Here the
$\Phi_{\gamma}^1$ probe is sensitive to all of the parameter space
shown and so its limit contour does not appear.}
\label{fig:reach50}
\end{figure}
Also noteworthy is the complementarity of traditional particle physics
searches and indirect dark matter searches.  Collider searches
require, of course, light superpartners.  High precision probes at low
energy also require light superpartners, as the virtual effects of
superpartners quickly decouple as they become heavy.  Thus, the LEP
and Tevatron reaches are confined to the lower left-hand corner, as
are, to a lesser extent, the searches for deviations in $B \to X_s
\gamma$ and $a_{\mu}$.  These bounds, and all others of this type, are
easily satisfied in the focus point models with large $m_0$, and
indeed this is one of the virtues of these models.  However, in the
focus point models, {\em all} of the indirect searches are maximally
sensitive, as the dark matter contains a significant Higgsino
component.  Direct dark matter probes share features with both
traditional and indirect searches, and have sensitivity in both
regions.  It is only by combining all of these experiments, that the
preferred region may be completely explored.\footnote{Note that the
complementarity referred to here is not the commonly recognized one,
which concerns the mass of the neutralino.  It is well-known that some
indirect searches are effective even for LSP masses in the TeV range,
well beyond the range of colliders.  However, such models are highly
unnatural, and they have not been considered here.}

\begin{figure}[!t]
\postscript{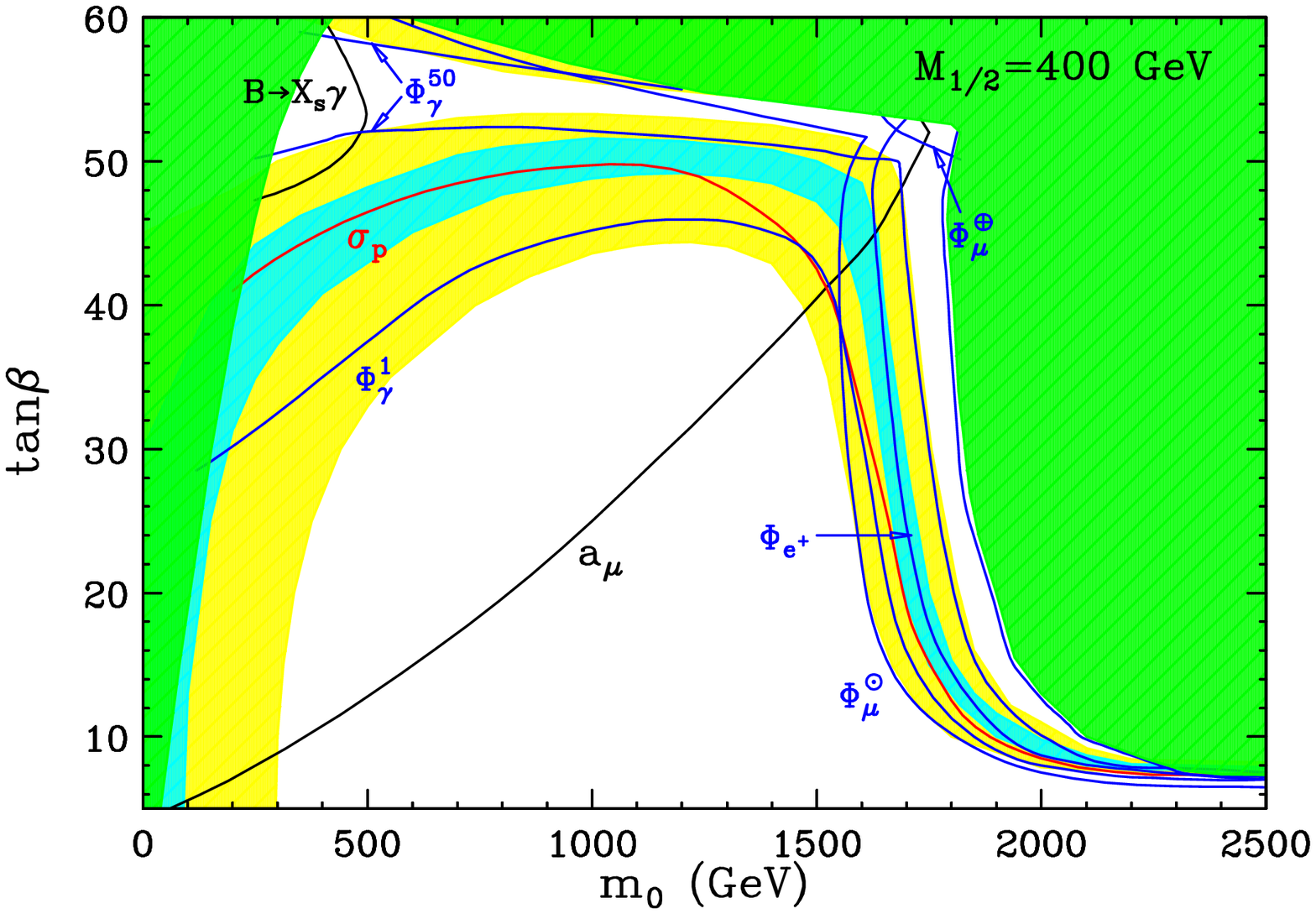}{0.74}
\caption{As in Fig.~\ref{fig:reach10}, but in the $(m_0, \tb)$ plane
for fixed $\mgaugino=400~\gev$, $A_0 = 0$, and $\mu > 0$.  The regions
probed are toward the green regions, except for $\Phi_{\gamma}^{50}$,
where it is between the two contours.  The top excluded region is
forbidden by limits on the CP-odd Higgs scalar mass.}
\label{fig:reach400}
\end{figure}
Finally, these results have implications for future colliders.  In the
cosmologically preferred regions of parameter space with $0.1 <
\Omegachi h^2 < 0.3$, all models with charginos or sleptons lighter
than 300 GeV will produce observable signals in at least one
experiment.  This is evident for $\tb=10$ and 50 in
Figs.~\ref{fig:reach10} and \ref{fig:reach50}.  In
Fig.~\ref{fig:reach400}, we vary $\tb$, fixing $\mgaugino$ to 400 GeV,
which roughly corresponds to 300 GeV charginos.  We see that the
preferred region is probed for any choice of $\tb$. (For extremely low
$\tb$ and $m_0$, there appears to be a region that is not probed.
However, this is excluded by current Higgs mass limits for $A_0 = 0$.
These limits might be evaded if $A_0$ is also tuned to some extreme
value, but in this case, top squark searches in Run II of the Tevatron
will provide an additional constraint.)

These results imply that if any superpartners are to be within reach
of a 500 GeV lepton collider, some hint of supersymmetry must be seen
before the LHC begins collecting data.  This conclusion is independent
of naturalness considerations.  While our quantitative analysis is
confined to minimal supergravity, we expect this result to be valid
more generally.  For moderate values of $\tb$, if the dark matter is
made up of neutralinos, they must either be light, Bino-like, or a
gaugino-Higgsino mixture.  If they are light, charginos will be
discovered.  If they are Bino-like, light sfermions are required to
mediate their annihilation, and there will be anomalies in low energy
precision measurements. And if they are a gaugino-Higgsino mixture, at
least one indirect dark matter search will see a signal.  For large
$\tb$, low energy probes become much more effective and again there is
sensitivity to all probe superpartner spectra with light
superpartners. Thus it appears, on qualitative grounds, that all
models in which the scalar masses are not widely separated, and the
charginos are not extravagantly heavy, will be accessible prior to LHC
operation.

\section{Conclusions}
\label{sec:conclusions}

In this paper, we have examined a wide variety of indirect dark matter
detection signals.  These include neutrinos from annihilation of dark
matter in the cores of the Earth and Sun, continuum gamma rays from
annihilation in the galactic center, and positron excesses in cosmic
rays from annihilation in the local solar neighborhood.  In each case,
the experimental landscape will be transformed in the next few years
by experiments that are running or being mounted.  We have tabulated
the salient features and reaches of some of the most promising
experiments in the previous sections.

We have evaluated the prospects for dark matter detection in the
framework of minimal supergravity.  This framework incorporates many
of the most compelling features of supersymmetry.  Previously, this
framework has been thought to predict a Bino-like LSP.  That severely
limited its utility for dark matter studies.  However, recent work has
made it clear that gaugino-Higgsino mixtures and even Higgsino-like
LSPs are also quite naturally realized in minimal supergravity.  We
have been careful to include the full range of possibilities, with
important (and positive) implications for future dark matter searches.

Let us note in passing that in our parametrization of experimental
probes, the case of no-scale supergravity~\cite{Lahanas:1987uc},
recently revived in the context of gaugino-mediated supersymmetry
breaking~\cite{Kaplan:2000ac,Chacko:2000mi,Schmaltz:2000gy,%
Schmaltz:2000ei}, can be regarded as the special case $m_0 = 0$.
Experimental probes of these models are simply evaluated by
restricting to the $m_0 = 0$ axis.  Several experiments, notably the
trilepton Tevatron search, direct dark matter searches, and Brookhaven
experiment E821 will have the power to confirm or exclude this
possibility in the near future.

We have concentrated here on discovery signals.  If a signal is
confirmed, precision measurements may allow experiments to determine
dark matter properties.  For example, as has been noted in the
literature, the neutralino's mass may be determined by the angular
spread of the signal in neutrino telescopes.  The energy spectrum of
gamma rays or positrons signals may provide similar information.  The
gaugino-ness may also be constrained; indeed, the existence of a
significant signal in itself would constitute evidence in favor of
mixed gaugino-Higgsinos.

The simplicity of minimal supergravity allows us to compare the
reaches of a great variety of probes.  We summarize by collecting
several of our main conclusions:

\begin{itemize}

\item Bino-like dark matter leads to suppressed rates for all indirect
dark matter signals.  In this case, unless the neutralino is extremely
light (near current bounds), all indirect signals are beyond detection
for the foreseeable future.

\item Higgsino-like dark matter cannot yield cosmologically
interesting relic densities in a straightforward way.  Studies that
assume Higgsino-like dark matter exaggerate the power of indirect
searches.

\item Mixed gaugino-Higgsino dark matter gives both relic densities in
the preferred range $0.1 \alt \Omegachi h^2 \alt 0.3$ and detectable
signals.  Such dark matter is naturally present in focus point models,
which are favored by low-energy constraints.

\item Experiments that are running or underway will transform the
prospects for indirect dark matter detection.  Among the most
promising experiments are the neutrino telescopes AMANDA, NESTOR, and
ANTARES; the gamma ray telescope MAGIC, and the satellite detector
GLAST; and AMS-02, the anti-particle/anti-matter search aboard the
International Space Station. For mixed gaugino-Higgsino dark matter,
these experiments will be sensitive to nearly all models with
cosmologically interesting neutralino relic densities, and are
competitive with next-generation direct search experiments, such as
CDMS (Soudan) and CRESST.

\item The various indirect searches rely on different sources of
neutralino annihilation (cores of the Earth or Sun, galactic center,
local solar neighborhood) and so are sensitive to different
assumptions.  In addition, some signals, particularly the continuum
photons, will be difficult to identify unambiguously as a dark matter
signal. Without actual data and detailed analyses, it is difficult to
make a more precise statement.  However, we have seen that many
experiments are sensitive to the same supersymmetric models, and given
the underlying uncertainties, redundancy is clearly a virtue.

\item Indirect searches are complementary to traditional particle
searches.  Separately they probe only portions of the cosmologically
preferred model space.  Combined, essentially all cosmologically
preferred models will produce at least a hint of a signal in one of
these experiments {\em before} the LHC begins operation.

\item In minimal supergravity models with $0.1 < \Omegachi h^2 < 0.3$,
{\em if there is no hint of supersymmetry before the LHC begins
operation, no superpartners will be within reach of a 500 GeV lepton
collider}.  Our arguments are independent of naturalness
considerations, and their qualitative structure suggests that similar
conclusions will remain valid in alternative frameworks.

\end{itemize}

\section*{Acknowledgments}

We thank P.~Blasi, L.~Hui, Z.~Ligeti, U.~Nierste and L.~Roszkowski
for helpful discussions and readings of the manuscript, and
W.~Hofmann, M.~Mori, I.~Moskalenko and T.~Weekes for correspondence.
This work was supported in part by the Department of
Energy under contracts DE--FG02--90ER40542 and DE--AC02--76CH03000,
and by the National Science Foundation under grant PHY--9513835.
J.L.F. acknowledges the support of a Frank and Peggy Taplin
Fellowship.

\end{document}